\documentclass[aps,prb,twocolumn,superscriptaddress,amsmath,amssymb]{revtex4-1}

\usepackage{physics}
\usepackage{graphicx}
\usepackage{xcolor}
\usepackage{amsmath}
\usepackage[utf8]{inputenc}
\usepackage{tikz}
\usepackage{bbm}

\newcommand{\up}{\uparrow}
\newcommand{\dn}{\downarrow}

\newcommand{\refeq}[1]{Eq.~(\ref{#1})}
\newcommand{\refsec}[1]{Sec.~\ref{#1}}
\newcommand{\reffig}[1]{Fig.~\ref{#1}}
\newcommand{\reftab}[1]{Tab.~\ref{#1}}

\newcommand{\pter}[1]{#1}

\newcommand{\cdag}{c^\dagger}

\begin{document}
\title{Exactly solvable model of strongly correlated $d$-wave superconductivity}
\author{Malte~Harland}
\affiliation{Institute of Theoretical Physics, University of Hamburg, Jungiusstra{\ss}e 9, 20355 Hamburg, Germany}
\author{Sergey~Brener}
\affiliation{Institute of Theoretical Physics, University of Hamburg, Jungiusstra{\ss}e 9, 20355 Hamburg, Germany}
\author{Mikhail~I.~Katsnelson}
\affiliation{Institute for Molecules and Materials, Radboud University, 6525AJ, Nijmegen, the Netherlands}
\author{Alexander~I.~Lichtenstein}
\affiliation{Institute of Theoretical Physics, University of Hamburg, Jungiusstra{\ss}e 9, 20355 Hamburg, Germany}
\date{\today}

\begin{abstract}
  We present an infinite-dimensional lattice of two-by-two plaquettes, the quadruple Bethe lattice, with Hubbard interaction and solve it exactly by means of the cluster dynamical mean-field theory. It exhibits a $d$-wave superconducting phase that is related to a highly degenerate point in the phase diagram of the isolated plaquette at that the groundstates of the particle number sectors $N=2,3,4$ cross. The superconducting gap is formed by the renormalized lower Slater peak of the correlated, hole-doped Mott insulator. We engineer parts of the interaction and find that pair hoppings between $X/Y$-momenta are the main two-particle correlations of the superconducting phase. The suppression of the superconductivity in the overdoped regime is caused by the diminishing of pair hopping correlations and in the underdoped regime by charge blocking. The optimal doping is $\sim 0.15$ at which the underlying normal state shows a Lifshitz transition. The model allows for different intra- and inter-plaquette hoppings that we use to disentangle superconductivity from antiferromagnetism as the latter requires larger inter-plaquette hoppings.
\end{abstract}

\maketitle

\section{Introduction}
High-temperature superconductivity  in cuprates\cite{Bednorz1986} can persist at temperatures up to $T\sim 100 \mathrm{K}$, which makes an understanding of their pairing mechanism non-trivial in the context of conventional theory of superconductivity and very interesting for theoretical and practical purposes. The large transition temperature is not the only peculiar characteristic of such materials. They are unconventional also by their anisotropic superconducting gap\cite{Wollman1993}, small superfluid density\cite{Emery1995}, and competing orders\cite{Fradkin2015}. Cuprates share a common quasi two-dimensional structure of layered copper-oxide compounds that are insulating and become superconducting upon doping with charge carriers\cite{Damascelli2003}. The different compounds of that family share a $d$-wave character of superconducting gap and antiferromagnetic order in the undoped insulating state. Furthermore, at larger hole doping and temperatures above the critical temperature they exhibit a very incoherent metal behavior characterized in particular by a linear temperature dependence of the resistivity\cite{Gurvitch1987} and by a pseudogap formation\cite{Renner1998}.

The Cu atoms of the stacked Cu-planes form a square lattice. On their bonds are oxygen atoms whose $p$-orbitals mediate electronic transitions from one Cu $d$-orbital to its nearest neighbor's $d$-orbital. This process is modeled by effective $d-d$ hopping with the amplitude $t$ that competes with the local screened Coulomb repulsion $U$. Further hoppings also exist, but they are smaller in their amplitude. The bandwidth of the $d$-orbitals is comparable to the interaction energy $U$. The broadly accepted minimal model to account for these competing electronic effects is the Hubbard model\cite{Hubbard1963,Zhang1988,Dagotto1994}. Despite its simple appearance that model in two and three dimensions can be solved by approximations only, contrary to the one-dimensional case that is exactly solvable by Bethe Ansatz\cite{Lieb1968}.

A simple but powerful approximation is the dynamical mean-field theory (DMFT)\cite{Georges1996} which accounts only for the local correlations by including only the local self-energy from an effective impurity model. The DMFT provides an exact solution in the formal limit of infinite dimensions but is questionable for two dimensions (2D). Phenomena such as the Mott transition\cite{Imada1998} and itinerant antiferromagnetism\cite{Fleck1998} (Slater physics) are captured by the infinite-dimensional DMFT, but are severely overestimated in low dimensions.

The DMFT can be extended by restricting the self-energy not to a single site, but to a cluster of several sites. Hence, this extension is called cluster DMFT (CDMFT)\cite{Lichtenstein2000,Maier2005,Kotliar2006}. The generalization to clusters is not unique and still debated\cite{Biroli2004,Vucicevic2018}. Regardless of the particular choice of CDMFT-''flavor'' it was found that intersite correlations within the cluster are sufficient to obtain a symmetry-broken d-wave superconducting (dSC) state. The minimal cluster is the two-by-two cluster (plaquette), since dSC order is defined on the bonds according to $d_{x^2-y^2}$-wave symmetry\cite{Lichtenstein2000}. In 2D CDMFT is an approximation and long-range correlations beyond the cluster can be important for a correct description of the dSC state in cuprates\cite{Gull2010,Sakai2012}. Therefore in the dSC state CDMFT aims to describe only the local formation of Cooper pairs. For example, CDMFT gives coexisting antiferromagnetic (AFM) and dSC orders\cite{Lichtenstein2000}, whereas in cuprates these phases do not coexist. The reason is that CDMFT does not distinguish between long-range and short-range AFM order if the corresponding correlation length is much larger than the lattice constant. Thereby, it also neglects the stripe order phase of cuprates which has been found to be suppressed by the next-nearest neighbor hopping within the Hubbard model\cite{Zheng2017,Jiang2018}.

In this work we present a detailed analysis of the infinite dimensional quadruple Bethe lattice model within the CDMFT. Similar to the well-known Mott transition found in the simple Bethe lattice\cite{Georges1996,Rozenberg1999,Bulla2001,Eckstein2005} and the correlated Peierls insulator transition in the double Bethe lattice\cite{Moeller1999,Hafermann2009,Najera2018}, we find the dSC transition in the strong-coupling quadruple Bethe lattice\cite{Harland2016}. This choice of setup is complementary to prior studies in the sense that we investigate a less accurate model of infinite dimensionality but in return obtain an exact solution. Compared to the simple Bethe lattice the local Hilbert-space size is increased, from $4$ of the Hubbard site to $256$ of the Hubbard plaquette. This opens up new degrees of freedom that can interact with the mean-field environment. In particular, we focus on those plaquette eigenstates\cite{Altman2002,Haule2007,Gull2008,Ferrero2009,Sordi2010}, that define the dSC and cross at a quantum critical point\cite{Harland2016} (QCP) of the plaquette. This point is particularly interesting as quantum critical behavior\cite{Mikelsons2009,Vidhyadhiraja2009,Vucicevic2018} has been found for the square lattice by CDMFT studies of the pseudogap phenomenon that has been suggested to originate from negative interference of hybridizing plaquette states\cite{Merino2014,Gunnarsson2015,Harland2016,Gunnarsson2018}.

In this paper we start with a presentation of the quadruple Bethe lattice and the single-particle basis we use in \refsec{sec:modelmethod}. In \refsec{sec:twobytwo} we provide an overview of the isolated cluster's Hilbert space, that is the auxiliary system of our CDMFT mapping. The opposite limit of non-interacting Bethe lattices is presented in \refsec{sec:nonint}. In \refsec{sec:umu} we analyse the dependence of the dSC order parameter on the screened Coulomb repulsion and the chemical potential for small Bethe lattice hoppings, i.e. plaquette hybridizations, for that the dSC order is dominant and other orders are less pronounced. In \refsec{sec:pkcoupling} we show how different components of the two-particle interaction promote or interfere with the dSC order. Dynamical properties, such as quasiparticle characterization and spectral functions are presented in \refsec{sec:spectral}. Finally, larger Bethe lattice hoppings yield more pronounced antiferromagnetic order, see \refsec{sec:otherorders}, and an extended Bethe lattice hopping allows us to tune the non-interacting density of states more similar to a van-Hove singularity, that is presented in \refsec{sec:tbnnn}.

\section{Model \& Method\label{sec:modelmethod}}
As stated above, the correlated $d$-electrons of the copper-oxide planes are described by the Hubbard model
\begin{equation}
  \label{eq:hubbard}
  H = \sum_{ij\sigma}t_{ij}c^{\dagger}_{i\sigma}c_{j\sigma}+U\sum_ic^{\dagger}_{i\up}c_{i\up}c^{\dagger}_{i\dn}c_{i\dn},
\end{equation}
with fermionic creation/annihilation operators $c^\dag$/$c$, sites $i,j$ and spins $\sigma$. It contains a hopping term $t_{i,j}$ that for lattice structures becomes diagonal in $k$-space and therefore promotes delocalization of the charge. Albeit, the quadruple Bethe lattice is only a pseudolattice in that regard since it does not exhibit translational invariance. But still, its sites are equivalent due to its self-similar structure. The screened local Coulomb repulsion $U$ is diagonal in site-space and promotes charge localization. The chemical potential $\mu$ can be written explicitly, or it can be absorbed into the diagonal, local part of $t_{ij}$.

\begin{figure}
  \includegraphics[scale=1]{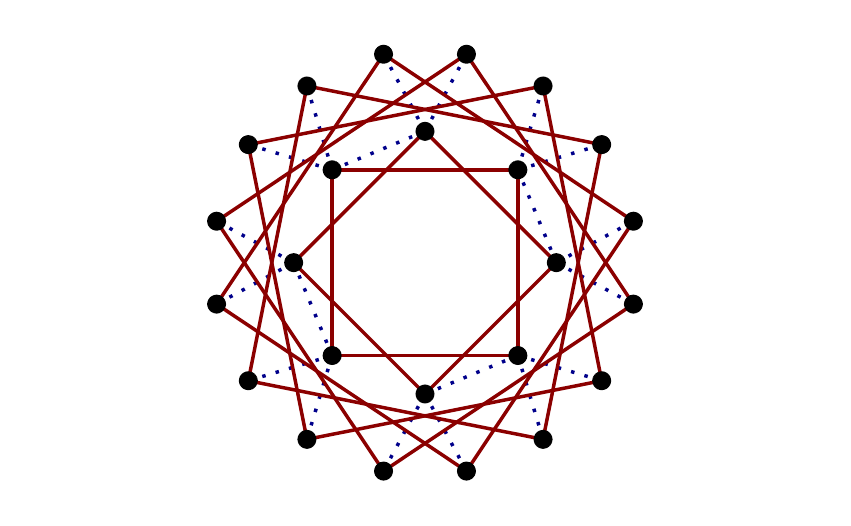}
  \caption{Quadruple Bethe lattice, four Bethe lattices (dotted lines) interconnected via plaquettes (solid lines). The coordination number for each Bethe lattice of this figure is set to $z=3$, and six sites of each Bethe lattice are depicted. An entire Bethe lattice exhibits an infinite number of sites with self-similar structure. Next-nearest neighbor hoppings of the plaquette are omitted for convenience.}
  \label{fig:structure}
\end{figure}
The quadruple Bethe lattice is constructed from four Bethe lattices, that are plaquette-wise connected, see \reffig{fig:structure}, i.e. equivalent sites of the four Bethe lattices form a two-by-two plaquette. The coordination number of the Bethe-lattices is set to $z=\infty$ corresponding to infinite dimensions. We introduce three types of hopping. The first hopping $t$ connects sites of the Bethe-lattice with equivalent points of two neighboring Bethe-lattices, i.e. within plaquettes. We use $t=-1$ throughout, and its absolute value defines our energy unit. The second hopping $t^\prime$ connects with equivalent points of the one remaining Bethe-lattice and thus occurs on the next-nearest neighbor bond of the plaquette. The third hopping $t_b$ connects sites within the Bethe-lattices, i.e. between plaquettes. We write the plaquette hopping matrix in site basis as
\begin{align}
  t^p =
  \begin{pmatrix}
    \begin{array}{rrrr}
      -\mu&t&t&t^\prime\\
      t&-\mu&t^\prime&t\\
      t&t^\prime&-\mu&t\\
      t^\prime&t&t&-\mu
    \end{array}
  \end{pmatrix}.
\end{align}
Then, we can decompose the kinetic energy $H_t$, the first term of \refeq{eq:hubbard}, into hopping within plaquettes and between plaquettes, i.e. within Bethe lattices
\begin{equation}
  \label{eq:ht}
  H_t = t_b \sum_{<r, r^\prime> R \sigma} c^\dag_{r^\prime R \sigma} c_{r R \sigma} + \sum_{r R R^\prime} t^p_{R R^\prime} c^\dag_{rR\sigma} c_{rR^\prime \sigma}.
\end{equation}
The original site label $i$ has been rewritten as a position within the Bethe lattice $r$ and position within the plaquette $R$. The summation over $<r, r^\prime>$ is performed over nearest neighbors. In principle the Bethe hopping can also be a matrix, but since we focus mostly on the case of a scalar $t_b$, we will restrict the following derivation to it. The generalization to a matrix Bethe hopping is straight forward, and we apply it in \refsec{sec:tbnnn} only.

\begin{figure}
  \begin{tikzpicture}
    \draw (0, 0) node[inner sep=0] {\includegraphics{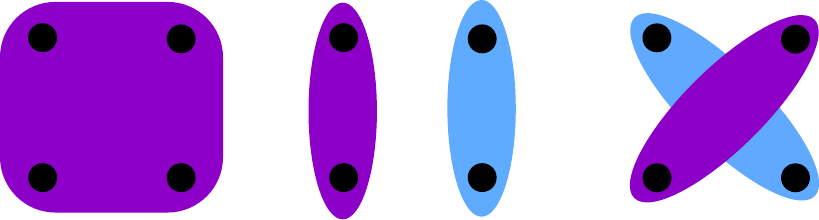}};
    \draw[anchor=north east] (-4+0.125, -1+0.125) node {\Large 0};
    \draw[anchor=north west] (-2-0.125, -1+0.125) node {\Large 1};
    \draw[anchor=south east] (-4+0.125, 1-0.125) node {\Large 2};
    \draw[anchor=south west] (-2-0.125, 1-0.125) node {\Large 3};
    \draw (-3, 1.5) node {\Large $\Gamma$};
    \draw (0, 1.5) node {\Large $X$};
    \draw (3.25, 1.5) node {\Large $M$};
  \end{tikzpicture}
  \caption{Illustration of the plaquette orbitals/momenta $\Gamma, X, M$ ($Y$ omitted), i.e. the basis that diagonalizes the hopping in plaquette-site space ($0,1,2,3$). Colors denote symmetries of the orbitals.}
  \label{fig:plaquetteorbitals}
\end{figure}
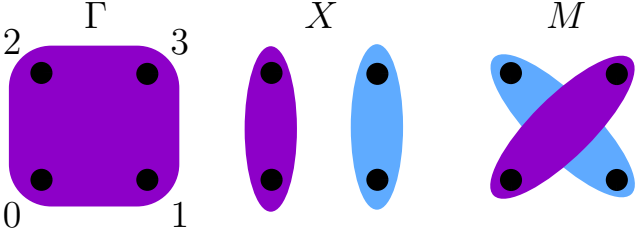
We apply a discrete Fourier transform to diagonalize the hopping of the plaquette from site space into the plaquette-momentum basis of momenta $K$ which can take four possible values, $\Gamma, M, X, Y$. The transformation applied to plaquette-site space reads
\begin{align}
  \label{eq:ktransform}
  \begin{split}
    T&=\frac{1}{2}
    \begin{pmatrix}
      e^{i\pter{\Gamma} \pter{R_0}}\ldots e^{i\pter{\Gamma} \pter{R_3}}\\
      e^{i \pter{M} \pter{R_0}}\ldots e^{i \pter{M} \pter{R_3}}\\
      e^{i \pter{X} \pter{R_0}}\ldots e^{i \pter{X} \pter{R_3}}\\
      e^{i \pter{Y} \pter{R_0}}\ldots e^{i \pter{Y} \pter{R_3}}\\
    \end{pmatrix}
    \\
    &= \frac{1}{2}
    \begin{pmatrix}
      \begin{array}{rrrr}
        1&1&1&1\\
        1&-1&-1&1\\
        1&-1&1&-1\\
        1&1&-1&-1\\
      \end{array}
    \end{pmatrix}
  \end{split}
\end{align}
with
\begin{align}
  \label{eq:ktransform2}
  \begin{split}
    \left(\pter{R}_0 ,..., \pter{R}_3\right) &=
    \begin{pmatrix}
      0 & 1 & 0 & 1\\
      0 & 0 & 1 & 1
    \end{pmatrix}a,\\
    \left( \pter{\Gamma}, \pter{M} , \pter{X}, \pter{Y} \right) &=
    \begin{pmatrix}
      0 & 1 & 1 & 0\\
      0 & 1 & 0 & 1
    \end{pmatrix}\frac{\pi}{a}\\
  \end{split}
\end{align}
with a unit length $a$. Due to the symmetry of the site-space, we can diagonalize the quadratic parts of the Hamiltonian $H$ in plaquette-momentum space. At this point the quadruple Bethe lattice can be regarded as a multiorbital simple Bethe lattice, see \reffig{fig:plaquetteorbitals}.

The interaction has to be transformed to $K$ basis as well. From the fact that it is local in site basis, one can already expect many terms in the plaquette-momentum basis. We apply the rank-2 tensor transformation $U \mapsto TTUT^\dag T^\dag$ also using plaquette-momentum conservation and obtain
\begin{align}
  \label{eq:utransf}
  \begin{split}
    H_U =\sum_{r K_1\dots K_4} U_{K_1\dots K_4}\cdag_{r K_1 \up}\, c_{r K_2 \up}\, \cdag_{r K_3 \dn}\, c_{r K_4 \dn}\\
  \end{split}
\end{align}
with the Hubbard interaction tensor $U_{K_1\dots K_4} = U \delta^{(2\pi/a)}_{K_1+K_3, K_2+K_4} /4$, where $\delta^{(2\pi/a)}_{K_1,K_2}=1$, when $K_1-K_2=2\pi n/a$ with integer $n$ and $=0$ otherwise. Depending on the relative values of the four plaquette momenta $K_1\dots K_4$ we can classify the terms of $U_{K_1\dots K_4}$ into intra-orbital ($\times 4$, e.g. $U_{XXXX}$), inter-orbital ($\times 12$, e.g. $U_{XXYY}$), pair-hopping ($\times 12$, e.g. $U_{XYXY}$), spin-flip ($\times 12$, e.g. $U_{XYYX}$), and correlated hopping ($\times24$ with all four momenta pairwise distinct). 

Regarding the superconducting order we use the Nambu spinor basis for its description
\begin{equation}
  \label{eq:basis}
  \tilde{c}^\dagger_r = \left( \cdag_{r\Gamma \up}\, c_{r\Gamma \dn}\, \cdag_{r M \up}\, c_{r M \dn}\, \cdag_{r X \up}\, c_{r X \dn}\, \cdag_{r Y \up}\, c_{r Y \dn} \right).
\end{equation}
It can be constructed efficiently by particle-hole transforming the spin-$\dn$ part of the conventional spinor representation. This is sufficient if no other than $S_z=0$ spin-structures are considered for the pairing. Next, we address the Hubbard-Hamiltonian in Nambu basis. We apply the Nambu spinor construction and examine how $t_{ij}$, $\mu$ and $U$ transform. This is done by using the anticommutation rules. We obtain
\begin{align}
  \label{eq:nambuhubbard}
  \begin{split}
    \tilde{t}^p_{\sigma}&=t^p\left(\delta_{\sigma \up}-\delta_{\sigma \dn}\right),\\
    \tilde{\mu}_\sigma&=\left(\mu+U\right)\delta_{\sigma \up}-\mu\delta_{\sigma \dn},\\
    \tilde{U}&=-U,
  \end{split}
\end{align}
where $\up$ and $\dn$ denote the indices of the Nambu spinor entries, i.e. spin-$\up$ particles and spin-$\dn$ holes. The Bethe hopping $t_b$ transforms under the Nambu spinor construction in the same way as the plaquette hopping $t^p$.

We use the CDMFT\cite{Georges1996,Lichtenstein2000,Maier2000,Maier2005,Kotliar2006} to map the lattice problem to the Anderson impurity model of an impurity, i.e. the cluster, with (quartic) interaction and a bath of non-interacting, but potentially renormalized, particles. The environment is defined by the dynamical mean-field (bath Green function) $\mathcal{G}(i\omega_n)$. In particular for the quadruple Bethe lattice with infinite coordination of the Bethe lattices, the CDMFT becomes exact as the self-energy
\begin{equation}
  \label{eq:dyson}
  \Sigma(i\omega_n) = \mathcal{G}^{-1}(i\omega_n) - G^{-1}(i\omega_n)
\end{equation}
exists only within the plaquettes and not between them. \refeq{eq:dyson} is the Dyson equation and relates $\Sigma$ to the local Green function $G$ and the bath Green function $\mathcal{G}$. They depend on Matsubara frequencies $\omega_n = \pi(2n+1)/\beta$ with inverse temperature $\beta$. In this study we use for the quadruple Bethe lattice calculations $\beta = 100$ throughout (though for calculations on the isolated plaquette we also use $\beta = 30$). The self-consistency condition to treat AFM and dSC order reads
\begin{widetext}
\begin{gather}
  \label{eq:selfconsistency}
  \begin{split}
    G^{-1}_{A K \sigma, A K^\prime \sigma^{\prime}}(i\omega_n) = i\omega_n \delta_{K K^\prime} \delta_{\sigma \sigma^\prime}+(\mu\delta_{K K^\prime}-t^p_{K K^\prime})\sigma^z_{\sigma \sigma^\prime}
    -t_{b}^2 \sum_{\tau \tau^\prime} \sigma^z_{\sigma \tau}G_{B K\tau, B K^\prime \tau^{\prime}}(i\omega_n)\sigma^z_{\tau^\prime \sigma^\prime} - \Sigma_{A K \sigma, K^\prime j \sigma^{\prime}}(i\omega_n).
  \end{split}
\end{gather}
\end{widetext}
$K,K^\prime$ are labels for the plaquette momenta. $\sigma, \sigma^\prime, \tau, \tau^\prime$ label the Nambu-space, i.e. spin-$\up$ electrons or spin-$\dn$ holes. The Nambu representation also requires a transformation of the single-particle energies (\refeq{eq:nambuhubbard}), the chemical potential $\mu$, the matrix of plaquette-hoppings $t^p$ and the scalar Bethe hopping $t_b$. For that reason the third Pauli matrix $\sigma^z$ appears in \refeq{eq:selfconsistency}. 

The fact that the  Bethe lattice is bipartite allows us to additionally consider the possibilty of AFM symmetry breaking. We can divide the lattice into two sublattices of which we know how to transform their local Green functions into each other analytically. We describe the AFM of the sublattices $A$ and $B$ with Nambu-Green functions as
\begin{align}
  \label{eq:gagb}
  \begin{split}
    G_{B K \sigma, B K^\prime \sigma^\prime}(i\omega_n) = - \sum_{\tau \tau^\prime} R_{\sigma \tau}\, G^\ast_{A K \tau, A K^\prime \tau^\prime}(i\omega_n)\, R_{\tau^\prime \sigma^\prime}^\dag
  \end{split}
\end{align}
with the rotation matrix
\begin{equation}
  \label{eq:rotationmatrix}
  R = e^{i\pi \sigma^y /2}.
\end{equation}
\refeq{eq:gagb} describes a spin-flip accompanied by a particle-hole transformation due to the Nambu spinor formalism. For the diagonal entries of the Green function there is no difference in using the first ($\sigma_x$) or the second ($\sigma_y$) Pauli matrix for the rotation of \refeq{eq:rotationmatrix}. In contrast, off-diagonal (anomalous) entries obtain an additional minus sign from $\sigma_y$. We use this Berry phase in order not to change the dSC order for the Bethe sublattices $A$ and $B$. $\sigma_x$ would change the dSC according to an $X$/$Y$-flip. Thus, \refeq{eq:gagb} defines staggered spin, but homogenous dSC order.

The spin order within the plaquette can still be diverse for different solutions. $A/B$ sublattices not only support AFM order, but also a spin order that is ferromagnetic within the plaquette and antiferromagnetic with respect to the Bethe sublattices $A/B$. We will refer to the latter as plaquette antiferromagnetism (PAFM).

The dynamical mean-field is constructed as
\begin{align}
  \label{eq:gweiss}
  \begin{split}
    \mathcal{G}^{-1}_{A K \sigma, A K^\prime \sigma^{\prime}}(i\omega_n) &= i\omega_n \delta_{K K^\prime} \delta_{\sigma \sigma^\prime}+(\mu\delta_{K K^\prime}-t^p_{K K^\prime})\sigma^z_{\sigma \sigma^\prime}\\
    &-t_{b}^2 \sum_{\tau \tau^\prime} \sigma^z_{\sigma \tau}G_{B K\tau, B K^\prime \tau^{\prime}}(i\omega_n)\sigma^z_{\tau^\prime \sigma^\prime}.
  \end{split}
\end{align}
Together with the local interaction, $\mathcal{G}$ defines the impurity setup. \refeq{eq:gweiss} shows that the mean-field of sublattice $A$ is constructed from the local properties of sublattice $B$. In the following we drop the Bethe lattice index $r = A/B$ for convenience. The numerical solution of the impurity Green function is obtained by the hybridization expansion continuous time quantum Monte-Carlo method\cite{Werner2006,Werner2006a,Gull2011,Parcollet2015,Seth2016} (CTHYB). The self-consistency is closed with the Dyson equation and by demanding that the local lattice Green function equals the impurity Green function which is inserted into the right-hand side of \refeq{eq:selfconsistency} until convergence is reached. In our implementation \refeq{eq:selfconsistency} is also used to iteratively find $\mu$ in the case of a certain filling is set as a parameter rather than $\mu$ directly.

The numerics can be implemented efficiently using symmetries and blockstructure of the Green function. For our setup the Matsubara-Green function has the structure
\begin{align}
  \label{eq:gblocks}
  &\hspace{20pt} \Gamma \hspace{33pt} M \hspace{41pt} X \hspace{38pt} Y \nonumber\\
  G =&
  \begin{pmatrix}
    \begin{array}{c|c||c|c}
      \gamma & 0 & a & 0\\ \hline
      0 & -\gamma^\ast & 0 & a^\ast\\ \hline \hline
      a & 0 & m & 0\\ \hline
      0 & a^\ast & 0 & -m^\ast\\
    \end{array}
    & 0 \\
    0 &
    \begin{array}{c|c||c|c}
      x & -d & \tilde{a} & \pi\\ \hline
      -d & -x^\ast & -\pi & \tilde{a}^\ast\\ \hline \hline
      \tilde{a} & -\pi & y & d\\ \hline
      \pi & \tilde{a}^\ast & d & -y^\ast\\
    \end{array}
  \end{pmatrix}.
\end{align}
It contains the two-by-two Nambu blocks of spin-$\up$ particles and spin-$\dn$ holes and additionally four-by-four blocks in plaquette momentum basis, the $\Gamma M$- and the $XY$-blocks. $d$ and $a$/$\tilde{a}$ stand for dSC and AFM orders, respectively. AFM breaks the plaquette point-group symmetry in such a way, that $\Gamma$, $M$ and $X$,$Y$ are pairwise coupled. In the plaquette momentum basis AFM order is reflected by non-zero $a$/$\tilde{a}$ off-diagonals. Furthermore, dSC order breaks the plaquette symmetries so that the $X$-$Y$ degeneracy is lifted, but off-diagonals are introduced only in Nambu-space, $d$ and $-d$. The diagonal-part of the $X$/$Y$-block is not affected by the dSC symmetry breaking, and thus $y=x$. The entries of $\pi$ describe spin-triplet superconductivity $\pi SC$ which we study in \refsec{sec:otherorders}. The anomalous part of the Green function has non-zero elements only in the $XY$-block. It can be written as
\begin{equation}
  \label{eq:ganom}
  F =
  \begin{pmatrix}
    \begin{array}{rr}
      -d & \pi\\
      -\pi & d
    \end{array}
  \end{pmatrix},
\end{equation}
for that the entries of $\pi$ show the symmetry $F_{XY}^{\up \dn} = - F_{YX}^{\up \dn} = F_{XY}^{\dn \up}$ and hence also the spin-triplet pairing. In contrast $d = F^{\up\dn}_{XX} = - F^{\dn\up}_{XX}$ which is a spin-singlet structure. Note, that in the present study non-zero entries for $\pi$ occur only simultaneously with the coexistence of dSC and AFM.

DMFT calculations of broken symmetries can be done efficiently by introducing seeds with the proper symmetry for the first DMFT-iteration and subsequently running additional loops until convergence. For example regarding dSC, we initialize the anomalous Green function with
\begin{align}
  \label{eq:anomiwinit}
  d^{\mathrm{init}}(i\omega_n) &= \frac{d_0\beta}{2}\left( \delta_{n,-1} + \delta_{n, 0}\right)
\end{align}
for some small $d_0$. This function transforms into a cosine in imaginary time that is symmetric and real.

\section{Two-by-two plaquette\label{sec:twobytwo}}
The low-energy many-body states of the Hubbard two-by-two plaquette, around $\expval{N} = 3$ filling, have been considered as an essential element of the description of superconductivity in cuprates. Prior investigations have shown\cite{Altman2002,Haule2007,Harland2016}, that the relevant low-energy subspace of the $256$ plaquette-states contains $6$ states: a $N=2$ spin-singlet with the symmetry of the plaquette-$\Gamma$ orbital $\ket{2, 0,\Gamma}$, two $N=3$ spin-doublets with $X$/$Y$ symmetries $\ket{3, \frac{1}{2}, \frac{X}{Y}}$ and a $N = 4$ spin-singlet of $\Gamma$ symmetry $\ket{4, 0,\Gamma}$. Note, that we use the notation of $\ket{3, \frac{1}{2}, \frac{X}{Y}}$ for the sector of the four degenerate states. In addition to these most important 6 states there are also a $N=4$ spin-triplet $\ket{4,1,M}$ and a $N = 3$ spin-quadruplet $\ket{3, \frac{3}{2}, M}$, that become important for large $U$ ($t-J$-limit). In this section we use $t^\prime = 0.3$. Calculations for $t^\prime = 0$ show qualitatively similar results although the QCP is shifted to larger values of $\mu$ and $U$.

\begin{figure}
  \includegraphics{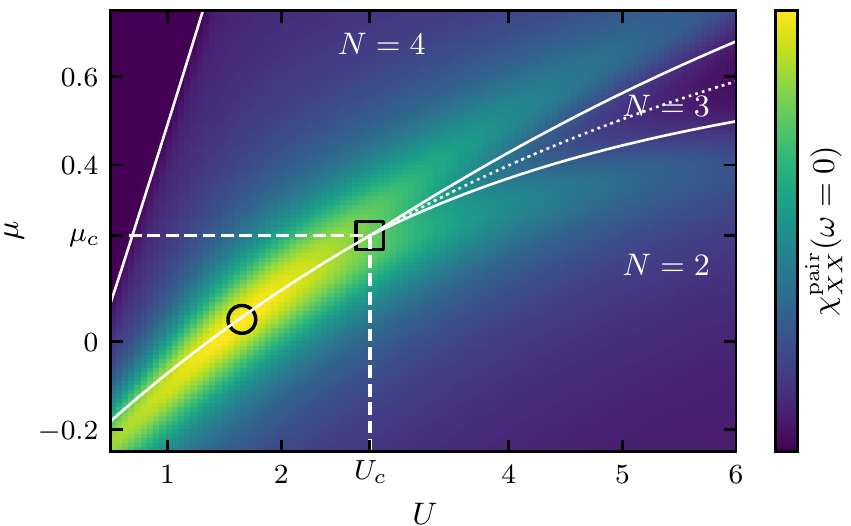}
  \caption{Retarded pairing susceptibility  $\chi^{\mathrm{pair}}(\omega)$ of pairs with plaquette momentum $X$ in the isolated plaquette dependent on the screened Coulomb repulsion $U$ and chemical potential $\mu$. The groundstate sectors $N=2,3,4$ (solid lines) cross at a quantum critical point (square) with $U_c = 2.78$ and $\mu_c = 0.24$. The maximum of $\chi^{\mathrm{pair}}(\omega)$ (black circle) lies at the $N=2,4$ crossover that becomes a non-groundstate crossover at $U_c<U$ (dotted line).}
  \label{fig:chixxpair}
\end{figure}
The instability towards dSC order can be observed already in the isolated plaquette using exact diagonalization. The pairing susceptibility
\begin{align}
  \label{eq:chipair}
    \chi^{\mathrm{pair}}_{XX}(\tau) = \expval{\mathrm{T}_\tau c_{X\up}(\tau) c_{X\dn}(\tau) c^\dag_{X\dn}(0) c^\dag_{X\up}(0)},  
\end{align}
with imaginary time ($\tau$) ordering operator $\mathrm{T}_\tau$ can be calculated using the Lehmann representation. The retarded pairing susceptibility at Fermi level $\chi^{\mathrm{pair}}(\omega = 0)$ shows large values in the $\mu$/$U$-phase diagram at the boundary of $N=2,4$, see \reffig{fig:chixxpair}. In the $N=4$ sector, where  $\ket{4,0,\Gamma}$ is the groundstate, $\ket{2,0,\Gamma}$ describes a bosonic two-hole excitation\cite{Altman2002}. $\chi^{\mathrm{pair}}_{XX}$ has its maximum close to $U=2$. Moreover, in this phase diagram the quantum critical point where 6 many-body states of the sectors $N=2,3,4$ cross can be seen at $U_c = 2.78$ and $\mu_c = 0.24$.

\begin{figure}
  \includegraphics{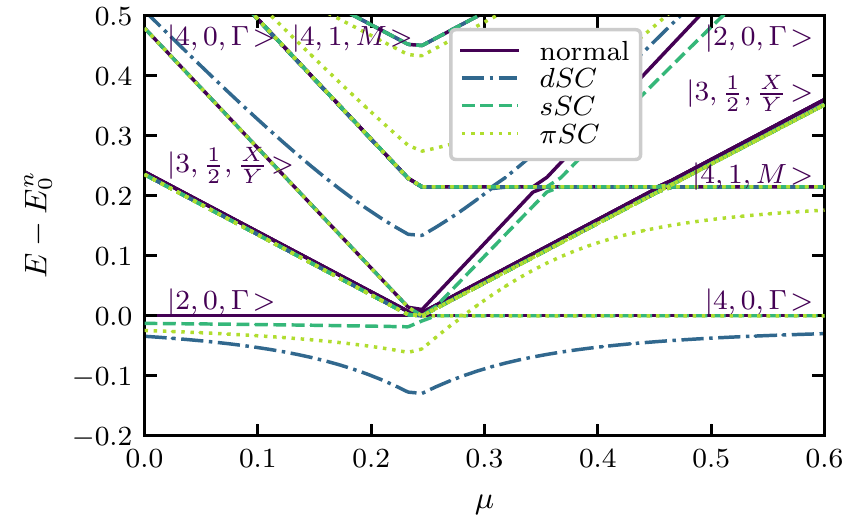}
  \caption{Plaquette eigenenergies $E$ as functions of the chemical potential $\mu$ around $\mu = 0.24, U=2.78$ at that $N=2,3,4$ cross. All energies are plotted relative to the groundstate-energy of the normal state $E^n_0$. The kets label the normal states particle number $N$, spin number $S$ and plaquette momentum $K$. The superconducting ($SC$) fields are $\Delta_x =0.1$ for the different orders $x$: $d$-wave $dSC$, $s$-wave $sSC$ and spin-triplet $\pi SC$.}
  \label{fig:e_mu}
\end{figure}
To get a view on the low-energy subspace of the plaquette we present in \reffig{fig:e_mu} the energy dependence of the states as a function of $\mu$ for constant $U=U_c$. Additionally, we add different $N$-symmetry breaking fields
\begin{align}
  \label{eq:dscfield}
  \begin{split}
    h_{dSC} &= \Delta_{dSC} (c_{\up X} c_{\dn X} - c_{\up Y} c_{\dn Y}) + \mathrm{h.c.},\\
    h_{sSC} &= \Delta_{sSC} (c_{\up X} c_{\dn X} + c_{\up Y} c_{\dn Y}) + \mathrm{h.c.},\\
    h_{\pi SC} &= \Delta_{\pi SC} (c_{\up X} c_{\dn Y} + c_{\dn X} c_{\up Y}) + \mathrm{h.c.}
  \end{split}
\end{align}
of spin-singlet $s$-wave (sSC), spin-singlet $d$-wave (dSC) and spin-triplet ($\pi$SC) symmetries. The groundstate energy lowering by the $dSC$ order is the largest at the critical point $\mu_c \sim 0.24$, see \reffig{fig:e_mu}. Different absolute values of the slopes in \reffig{fig:e_mu} correspond to different particle number sectors of the normal state. The small-$\mu$ and large-$\mu$ part have $\ket{2, 0, \Gamma}$ and $\ket{4, 0, \Gamma}$ as groundstates, respectively. The $dSC$-groundstate is a superposition of mainly these two and there crossing is avoided by the symmetry breaking. A contribution of $N=3$ to the $dSC$ groundstate is excluded since Cooper-pairs contain two electrons and therefore the groundstate has even parity, i.e. it is a superposition of particle number sectors of even particle numbers. 

Regarding $\Delta_{sSC}$, only the $\ket{2, 0, \Gamma}$ is lowered in energy, but not due to mixing with the low-energy $\ket{4, 0, \Gamma}$ of the normal state as this one is unaffected. For the $\pi SC$ field, the degeneracy of the spin-triplet $\ket{4,1,M}$ is lifted as only the $S^z=0$-state mixes with $\ket{2,0,\Gamma}$. The splitting of the two is visible, whereas one state is lowered in energy, the other is increased relative to the corresponding normal state. Without field, i.e. in the normal state, $\ket{2,0,\Gamma}$ and $\ket{4,1,M}$ cross around $\mu=0.35$, this crossing is avoided in the $\pi SC$ state. Among the considered symmetry breakings, the energy splitting of low-energy states with $dSC$-field is the largest. It is noticeable, that the main instability in the many-body physics of the Hubbard two-by-two plaquette is towards dSC order as it lowers the energy the most. 

\begin{figure}
  \includegraphics{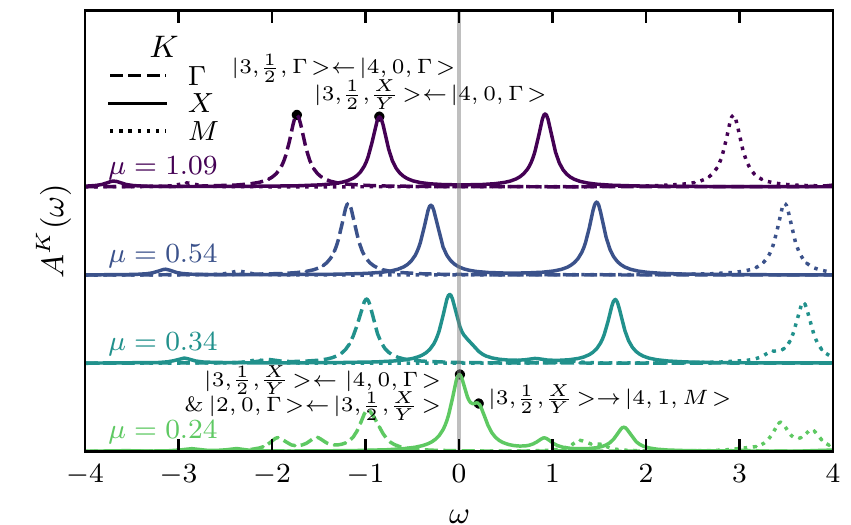}
  \caption{Plaquette momentum $K$ resolved Spectral function $A^K(\omega)$ of the isolated plaquette for different chemical potentials $\mu$. The peaks are identified with single-particle transitions of the plaquette eigenstates. $U = 2.78$, $t^\prime = 0.3$, $\beta = 30$ and Lorentzian broadening $\epsilon = \pi / \beta $.}
  \label{fig:ak}
\end{figure}
In addition to the transition of pairs we investigate also single-particle transitions of the isolated plaquette, see \reffig{fig:ak}. At half-filling ($\mu = 1.09$) we observe a four-peak structure of the spectral function. Since we are interested in hole-doping, we focus on the transitions below Fermi-level. The lowest has plaquette momentum $\Gamma$ and is a transition of the groundstate, $\ket{4,0,\Gamma} \rightarrow \ket{3,\frac{1}{2}, \Gamma}$. The second lowest and closest to Fermi level has plaquette momentum $X/Y$ and corresponds to $\ket{4,0,\Gamma} \rightarrow \ket{3,\frac{1}{2}, \frac{X}{Y}}$. Thus, $\ket{3,\frac{1}{2}, \frac{X}{Y}}$ describes the low-energy, one-particle excitation of plaquette momentum $X/Y$ of the $N=4$ system with groundstate $\ket{4,0,\Gamma}$.

Upon reducing $\mu$ the system gets hole-doped and the lower peak of $A^X(\omega)$ crosses Fermi-level. At $\mu=0.24$ is a groundstate crossover, the QCP, and therefore other transitions become active. At the QCP the transitions $\ket{4,0,\Gamma}\rightarrow \ket{3,\frac{1}{2}, \frac{X}{Y}}$ and $\ket{3,\frac{1}{2}, \frac{X}{Y}} \rightarrow \ket{2,0,\Gamma}$ occur on the same $\omega$. Furthermore, the peak has a pronounced shoulder from the transition $\ket{3,\frac{1}{2},\frac{X}{Y}} \rightarrow \ket{4,1,M}$. Thus in total around the QCP three prominent one-particle transitions exist close to Fermi level.

\section{Non-interacting quadruple Bethe lattice\label{sec:nonint}}
For the non-interacting case ($U=0$, $\Sigma = 0$) the Green function $G(i\omega_n)$ of \refeq{eq:selfconsistency} becomes the bare Green function $G^0(i\omega_n)$. Thus, we can solve \refeq{eq:selfconsistency} analytically and obtain
\begin{align}
  \label{eq:uzero}
  \begin{split}
    G_K^0(i\omega_n)&=\frac{2 \sigma_z}{\xi_K-\sqrt{\xi_K^2 - 4t^2_b}},\\
    \xi_K&=i\omega_n\sigma_z+\left(\mu - t_K^p\right)\mathbbm{1}.
  \end{split}
\end{align}
The third Pauli matrix $\sigma_z$ stems from the particle hole transformation of the hoppings and acts on Nambu space. The derivation of the analytical solutions depends on the fact, that all quantities can be diagonalized in spinor and $K$-space by a unitary transformation.

\begin{figure}
  \includegraphics{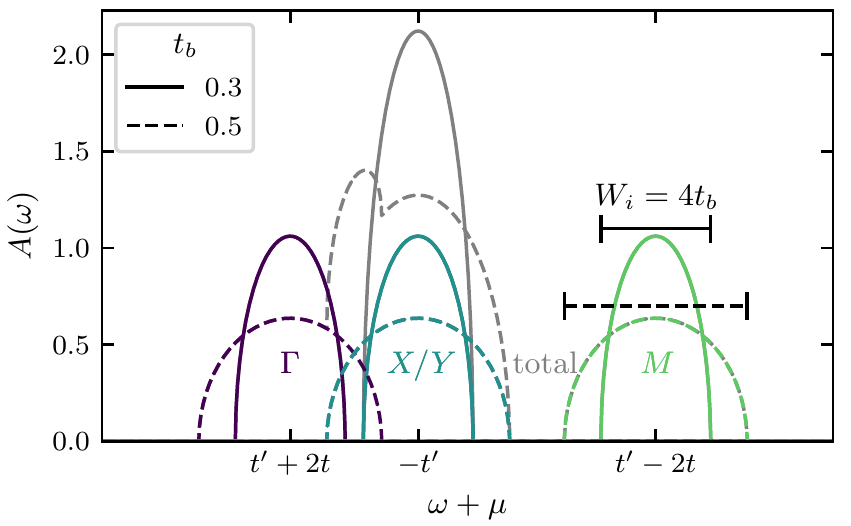}
  \caption{Semicircular densities of states of the non-interacting ($U=0$) quadruple Bethe lattice for the different orbitals/momenta and bethe-lattice hoppings $t_b$ ($t^\prime = 0.3$). The semicirculars have the same width $W_i$ for scalar $t_b$.}
  \label{fig:nonint}
\end{figure}
The spectral function corresponding to $G^0(i\omega_n)$ is shown in \reffig{fig:nonint}. It consists of four semicirculars of that two are degenerate corresponding to $X$ and $Y$. The semicirculars have a bandwidth of $W = 4t_b$ each. The positions of the semicirculars are defined by the eigenvalues of the hopping within the plaquette. Therefore, we have the lowest momentum/orbital $\Gamma$ at $\omega_\Gamma = t^\prime+2t -\mu$, the highest $M$ at $\omega_M = t^\prime-2t -\mu$ and $X/Y$ at $\omega_{X/Y} = -t^\prime -\mu$. The model is particle-hole symmetric for $t^\prime = 0$ and large values of $t^\prime$ or $t_b$ can make the orbitals overlap. 

\begin{figure}
  \includegraphics{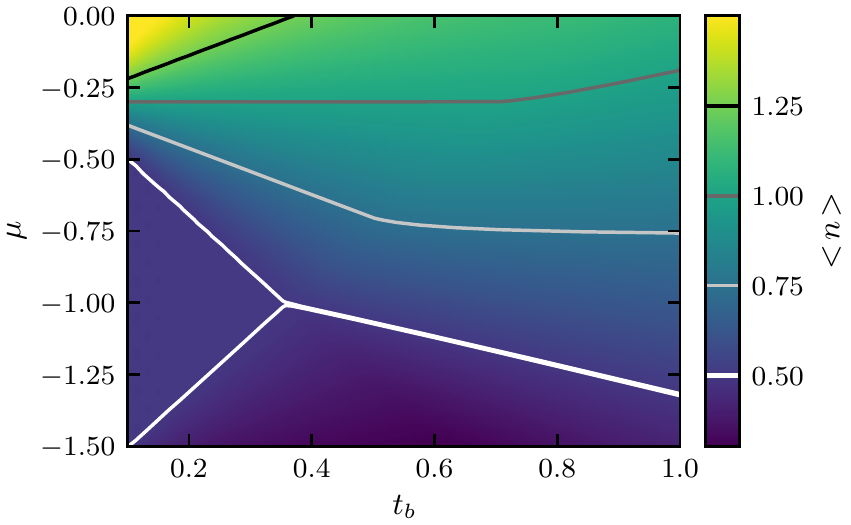}
  \caption{Filling $\expval{n}$ dependence on the chemical potential $\mu$ and the Bethe hopping $t_b$ for the non-interacting ($U=0$) quadruple Bethe lattice ($t^\prime = 0.3$).}
  \label{fig:nonintnmutb}
\end{figure}

The dependence of the filling on the chemical potential and the Bethe-hopping are shown in \reffig{fig:nonintnmutb}. In order to relate states of the isolated plaquette to solutions of the quadruple Bethe lattice, it can be useful to know the effect of $t_b$. From the non-interacting case we can learn how $t_b$ and $\mu$ change the filling. For small $t_b$ and $0.5 < \expval{n} < 1 $, $t_b$ reduces the particle occupation. There are mainly two effects that define this dependence. First, the semicircular at Fermi-level broadens, depending on whether its maximum is above or below Fermi-level it increases or decreases the filling. Second, an additional semicircular can broaden enough to also touch the Fermi-level and thereby change the filling.

\section{$\mu$-$U$ phase diagram\label{sec:umu}}
\begin{figure*}
  \includegraphics{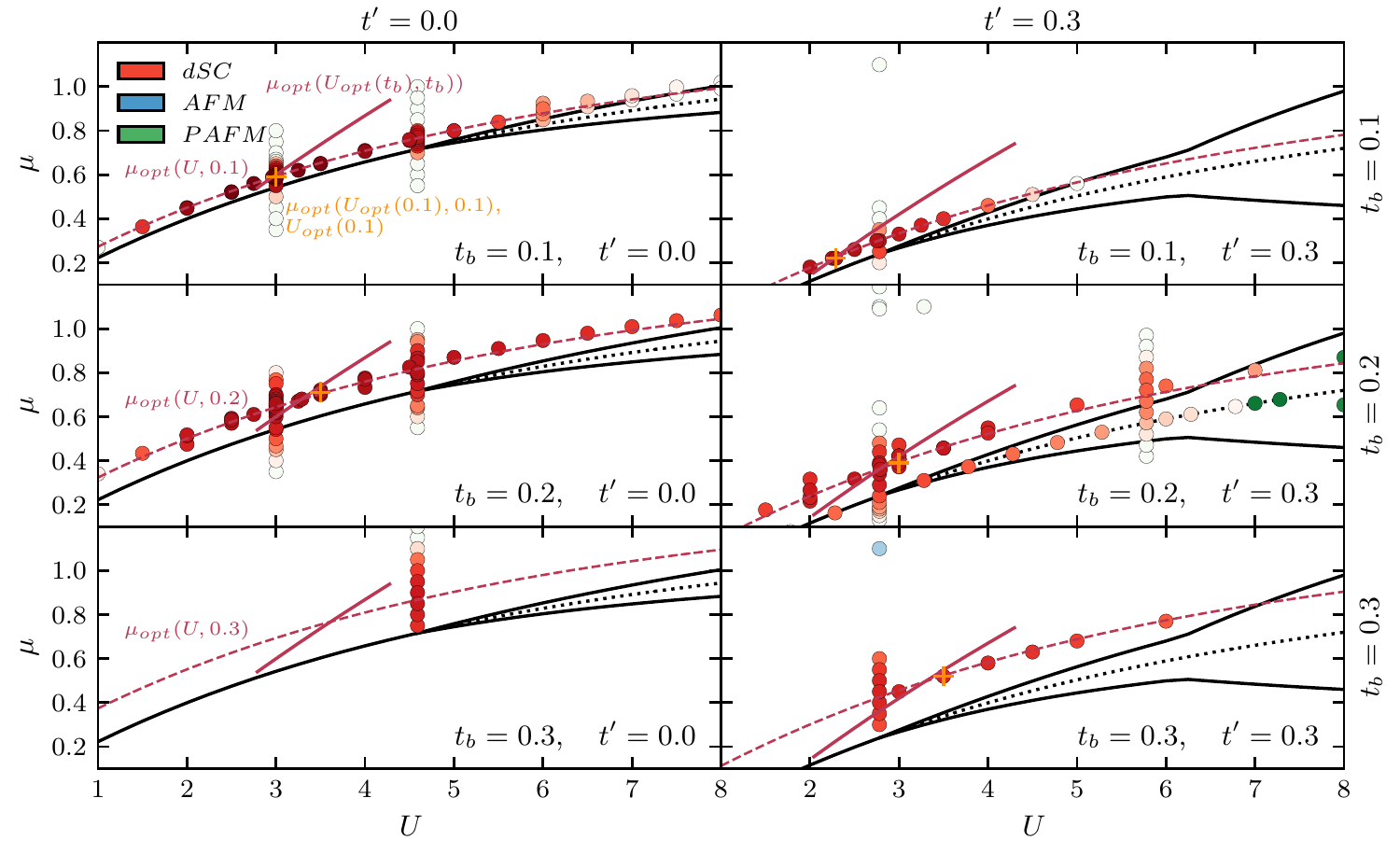}
  \caption{Phase diagrams of the quadruple Bethe lattice dependent on the chemical potential $\mu$ and Hubbard interaction $U$ for several next-nearest-neighbor hoppings $t^\prime$ and Bethe hoppings $t_b$. Considered spontaneously broken symmetries are $d$-wave superconductivity (dSC), antiferromagnetism (AFM) and plaquette antiferromagnetism (PAFM). The black lines denote the groundstate crossovers of the regions $N=2,3,4$ (bottom to top) of the isolated plaquette. The dotted black line marks the crossing of the $N=2,4$ sector-groundstates of the isolated plaquette. The maximum dSC order value per diagram is marked by ``+''. The dashed and solid colored lines correspond to $\mu_{opt}(U,t_b)$ and $U_{opt}(t_b)$ fits corresponding to $\Psi^{max}_{dSC}$, respectively.}
  \label{fig:phasediag_tb_tnnn}
\end{figure*}
In order to get an overview of the phases of the quadruple Bethe lattice and their relation to the states of the isolated plaquette, \reffig{fig:phasediag_tb_tnnn} presents several phase diagrams in the $\mu$-$U$-plane for different plaquette next-nearest-neighbor hopppings $t^\prime$ and Bethe hoppings $t_b$. The orders described by the selfconsistency condition of \refeq{eq:selfconsistency}, that exist for small $t_b$, are dSC, AFM and PAFM. Their order parameters are defined as
\begin{align}
  \label{eq:orderparameters}
  \Psi_{dSC} =& \frac{1}{4^2} \sum_{R R^\prime} \left( \cos \left[ \pter{X} \left( \pter{R}- \pter{R}^\prime\right)\right] - \cos \left[ \pter{Y} \left( \pter{R} - \pter{R}^\prime\right)\right]  \right)\nonumber \\
             &\times\expval{c_{R\up}c_{R^\prime\dn} - c_{R\dn}c_{R^\prime\up}},\nonumber\\
  \Psi_{AFM} =& \frac{1}{4} \sum_{R} e^{i \pter{M}\pter{R}} \expval{S^z_R},\\
  \Psi_{PAFM} =& \frac{1}{4} \sum_{R} e^{i \pter{\Gamma}\pter{R}} \expval{S^z_R},\nonumber
\end{align}
with the local spin along quantization axis $S^z_R = (n_{R\up}-n_{R\dn})/2$. By the symmetries of \refeq{eq:gblocks} and \refeq{eq:ganom} we obtain $\Psi_{dSC} = \Tr_{i\omega_n} F_{XX}(i\omega_n)$. The order parameters are calculated broad region around the QCP, where the groundstates $\ket{2,0,\Gamma}$, $\ket{3,\frac{1}{2},\frac{X}{Y}}$ and $\ket{4,0,\Gamma}$ cross. We stress that \reffig{fig:phasediag_tb_tnnn} combines information of two different systems, i.e. the phase boundaries of the isolated plaquette ($t_b=0$, $T=0$) and order parameters of the quadruple Bethe lattice ($t_b > 0$, $T=0.01$).

The most dominant order in that region for all $t^\prime$ and $t_b$ of \reffig{fig:phasediag_tb_tnnn} is the dSC. For small $t_b$ the dSC region is relatively narrow as a function of $\mu$. It broadens, and its maximum $\Psi^{max}_{dSC}$ decreases with increasing $t_b$, as if it is smeared. $t_b$ increases the width of the semicircular density of states keeping its area constant and thereby decreases its height. Thus, $t_b$ increases the energy window for fluctuations, i.e. more plaquette eigenstates from higher energies contribute to the solution of the quadruple Bethe lattice, but at the same time the amplitudes of the quantum fluctations can become smaller. This gives at least an intuition of $t_b$'s effect, the quantitative details are hidden in the CDMFT self-consistency.

The QCP of the plaquette shifts to smaller $\mu$ and smaller $U$ as $t^\prime$ is increased. For $t^\prime = 0.3$, we also find an additional crossover from the spin-doublet $\ket{3,\frac{1}{2},\frac{X}{Y}}$ to the spin-quadruplet $\ket{3, \frac{3}{2}, M}$, that is recognized by a kink in the phase boundaries around $U\sim 6$. In the quadruple Bethe lattice, at $t_b = 0.2$, in that region PAFM order is observed. It is spin-$3/2$ antiferromagnetism of plaquette ``supersites'', i.e. a quadruple Bethe lattice of ferromagnetic plaquettes and antiferromagnetic Bethe lattices. The cuprates show many competing orders near the dSC dome, such as stripes and spin/charge density waves, that have also been investigated in the framework of the Hubbard model or its limit, the $t$-$J$ model. However, it is unclear how the PAFM order found here could be related to those.

AFM is found for $t^\prime = 0.3, t_b = 0.3$ at large $\mu$, close to half-filling, with a relatively small order parameter, but in the considered parameters of \reffig{fig:phasediag_tb_tnnn} AFM is mostly absent. Heisenberg AFM is promoted by double occupations of sites that occur at half-filling. The effective spin exchange $J$ appears in the strong coupling regime of the Hubbard model, i.e. for large $U$\cite{Brinkman1970,Pruschke2003}. Therefore the predominant abscence of AFM within the phase diagrams of \reffig{fig:phasediag_tb_tnnn} seems reasonable as $t_b$ is small and $U$ has intermediate values. The fact that it appears only at $t^\prime = 0.3$ suggests that $t^\prime$ can cause an effectively enhanced $U$. A more detailed view on the AFM order will be provided below, in \refsec{sec:otherorders} where we discuss larger $t_b$.

\begin{figure}
  \includegraphics{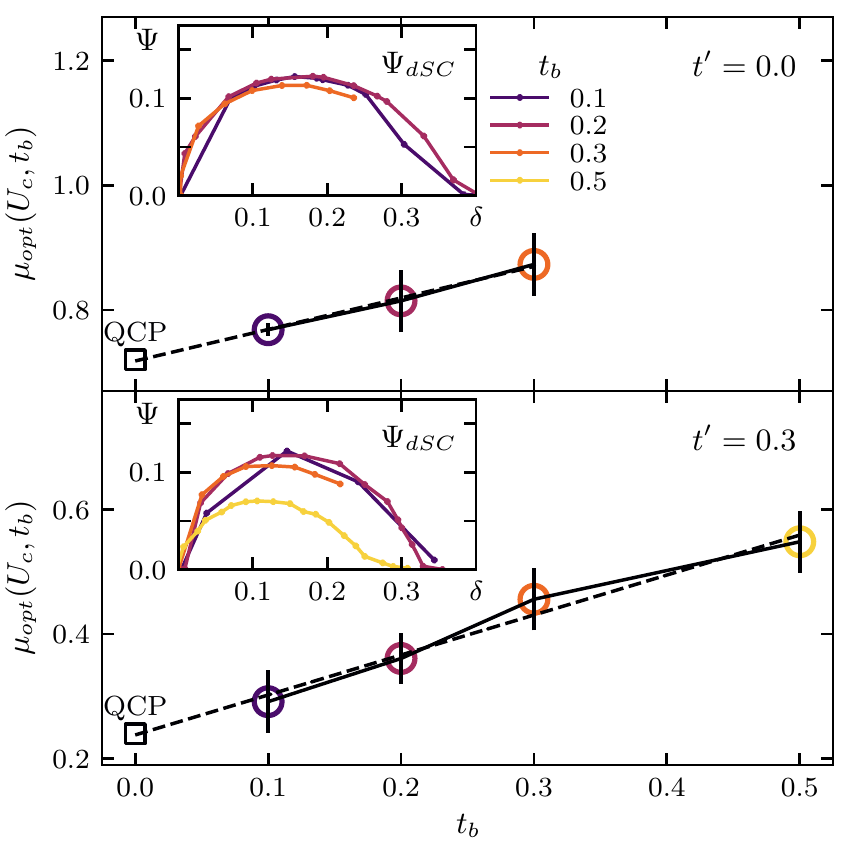}
  \caption{Chemical potential $\mu_{opt}$, that maximizes the $d$-wave superconducting order parameter $\Psi_{dSC}$ as a function of the Bethe hopping $t_b$, $U=U_c$. Linear fits (dashed) are performed for the next-nearest neighbor hoppings $t^\prime = 0$ (top) and $t^\prime = 0.3$ (bottom) separately and are extrapolated to $t_b=0$. The quantum critical point (QCP) of the isolated plaquette is shown, too. $\Psi_{dSC}$ is also shown as a function of the hole doping $\delta$ (insets) for different $t_b$ (color-coded).}
  \label{fig:muoptfitdscdop}
\end{figure}
In the following we locate and study the optimal parameter set $(\mu_{opt}, U_{opt}, t^{opt}_b)$ that corresponds to $\Psi^{max}_{dSC}$ using linear fits for fixed $t^\prime=0$ and $t^\prime = 0.3$. The optimal chemical potential $\mu_{opt}$, that corresponds to $\Psi^{max}_{dSC}$ as a function of $U$ is found on a line in $\mu$-$U$-plane that is parallel to the line $\mu_{24}$ of the plaquette's $\ket{2,0,\Gamma}$-$\ket{4,0,\Gamma}$-crossing, even if these two are not the groundstates. \reffig{fig:phasediag_tb_tnnn} shows this for small $0.1 \leq t_b \leq 0.3$. In \reffig{fig:nonintnmutb} we see a linear $t_b$-dependence of $\mu$ at constant filling. We write the linear model to fit the optimal chemical potential for a constant $t^\prime$
\begin{equation}
  \label{eq:muopt}
  \mu_{opt}(U, t_b) \simeq \mu_{24}(U) + \mu_{opt}^{(1)} t_b.
\end{equation}
$\mu_{24}(U)$ is calculated on the isolated plaquette and the coefficient of the linear shift by $t_b$, namely $\mu_{opt}^{(1)}$, is fitted to the numerical results of the quadruple Bethe lattice,
\begin{table}
  \begin{ruledtabular}
  \begin{tabular}{ccccc}
    $t^\prime$ & $\mu_{24}(U_c) = \mu_c$  & $\mu^{(1)}_{opt}$  & $U_{opt}^{(0)}$ & $U_{opt}^{(1)}$\\\hline
    $0$ & $0.72$ & $0.51\pm 0.02$ & $2.93$ & $1.79$\\
    $0.3$ & $0.24$ & $0.62\pm 0.05$ & $1.82$ & $5.04$\\
  \end{tabular}
  \end{ruledtabular}
  \caption{Fit-coefficients of the linear-$t_b$ models for the optimal chemical potential $\mu_{opt}$ and optimal Hubbard interaction $U_{opt}$ for different next-nearest-neighbor hoppings $t^\prime$. The offset of $\mu_{opt}$, i.e. $\mu_{24}$, is calculated in the isolated plaquette, it is the chemical potential at that $\ket{2,0,\Gamma}$ and $\ket{4,0,\Gamma}$ of the isolated plaquette cross.}  
  \label{tab:scopt}
\end{table}
see \reftab{tab:scopt} for the coefficients. The maxima in the doping-dependence of the dSC order parameter $\Psi_{dSC}(\delta)$ have been calculated via quadratic fits to the largest values. The data is presented in \reffig{fig:muoptfitdscdop} (insets). \reffig{fig:muoptfitdscdop} shows, that the $t_b$-dependence of the $\Psi^{max}_{dSC}$ is indeed linear. Furthermore, the extrapolation to $t_b=0$ points to $\mu_{24}$ of the isolated plaquette, that for $U=4.59$, $t^\prime=0$ and $U=2.78$, $t^\prime=0.3$ is the QCP. $\Psi^{max}_{dSC}$ at $t_b = 0.1$ is very similar for $t^\prime =0$ and $t^\prime =0.3$. For very small $t_b$ the quadruple Bethe lattice turns into isolated plaquettes and dSC vanishes.

So far, we have focused on a description in terms of energies and thus on $\mu$ rather than the observable hole doping $\delta$. In \reffig{fig:muoptfitdscdop} (insets) we present $\Psi_{dSC}$ depending on the doping. For small $t_b$ $t^\prime = 0$ and $t^\prime = 0.3$ share a maximum around $\delta \sim 0.15$, that is the optimal doping of cuprates\cite{Damascelli2003}. In particular, for the data of $t^\prime = 0.3$, at that we have also calculated solutions of $t_b = 0.5$, $\Psi^{max}_{dSC}$ shifts towards half-filling. It is remarkable, that the maximum at $\delta \sim 0.15$ is such a stable feature for different $t^\prime$ and $U$ at small $t_b \sim 0.1$, i.e. weakly hybridized plaquettes. Larger $t_b$ make the dSC dome results similar to 2D CDMFT studies at larger temperatures, where the dSC dome is closer to half-filling. In the 2D approximation of CDMFT the hybridization is solely determined by the self-consistency condition and there is no analogue to $t_b$. The present context can raise the question whether long-range correlations that are neglected by 2D CDMFT can effectively turn the system into more weakly hybridized plaquettes.

\begin{figure}
  \includegraphics{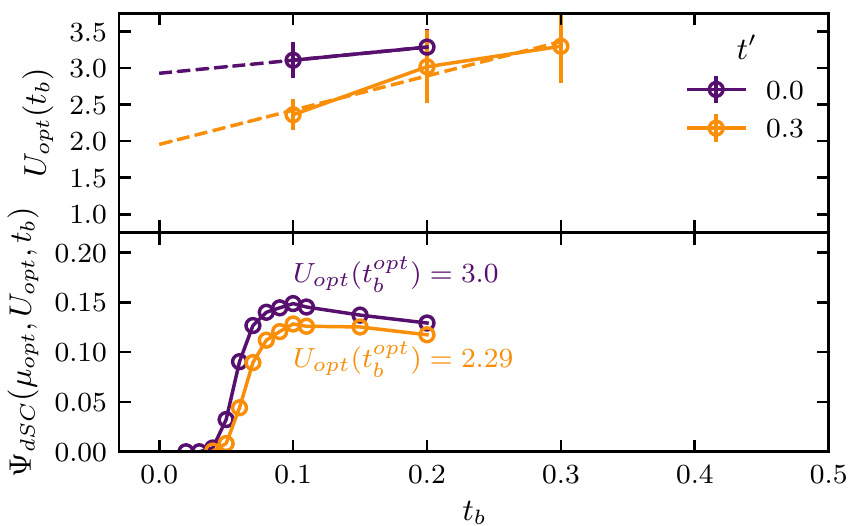}
  \caption{
    Top: Linear fit (dashed) of the optimal Hubbard interaction $U_{opt}$ as a function of the Bethe hopping $t_b$ at optimal doping ($\mu_{opt}$). The fit is performed for different next-nearest-neighbor hoppings $t^\prime$ separately.
    Bottom: $d$-wave superconducting order parameter along the optimal $\mu$-$U$-line as a function of $t_b$.}
  \label{fig:uoptscoopttb}
\end{figure}
With the fit of $\mu_{opt}(U, t_b)$ we can predict optimal doping, next we fit a linear model for optimal Hubbard interaction
\begin{equation}
  \label{eq:uopt}
  U_{opt}(t_b) \simeq U_{opt}^{(0)} + U_{opt}^{(1)} t_b,
\end{equation}
to find the optimal $U_{opt}(t_b)$ that maximizes $\Psi_{dSC}$ along the line described by $\mu_{opt}(U, t_b)$ in the $\mu-U$ phase diagram. But contrary to $\mu_{opt}(U,t_b)$, we need to fit the slope $U_{opt}^{(1)}$ and the offset $U_{opt}^{(0)}$. Furthermore, there is no motiviation from the non-interacting case as in the $\mu_{opt}$-fit. We use it only to estimate the position of the maximum $\Psi^{max}_{dSC}$ within the $\mu$-$U$ phase diagram, also for different $t_b$. \reffig{fig:uoptscoopttb} (top) shows the linear fit of $U_{opt}$, though only few points are taken into account. The fitted models predict the position, $(\mu_{opt}, U_{opt})$, of $\Psi^{max}_{dSC}$ dependent on $t_b$ in the $\mu$-$U$ plane, see \reffig{fig:phasediag_tb_tnnn}.

Along the line of $t_b$-dependent ($U_{opt}$, $\mu_{opt}$) the dSC order parameter exhibits a maximum at $t_b = 0.1$, see \reffig{fig:uoptscoopttb} (bottom), that is an order of magnitude larger than the temperature $T=0.01$ and smaller than the plaquette hopping $|t|=1$. The steep slope of $\Psi_{dSC}$ in \reffig{fig:uoptscoopttb} (bottom) at small $t_b$ is difficult to resolve accurately since the filling is very sensitive and small errors in the $\mu_{opt}$-estimate can cause strong noise. The steep slope is caused by the transition of the quadruple Bethe lattice into disconnected plaquettes. The $t_b$ dependence of $U_{opt}(t_b)$ is stronger for $t^\prime = 0.3$ than for $t^\prime = 0$ (\reftab{tab:scopt}).

\begin{figure}
  \includegraphics{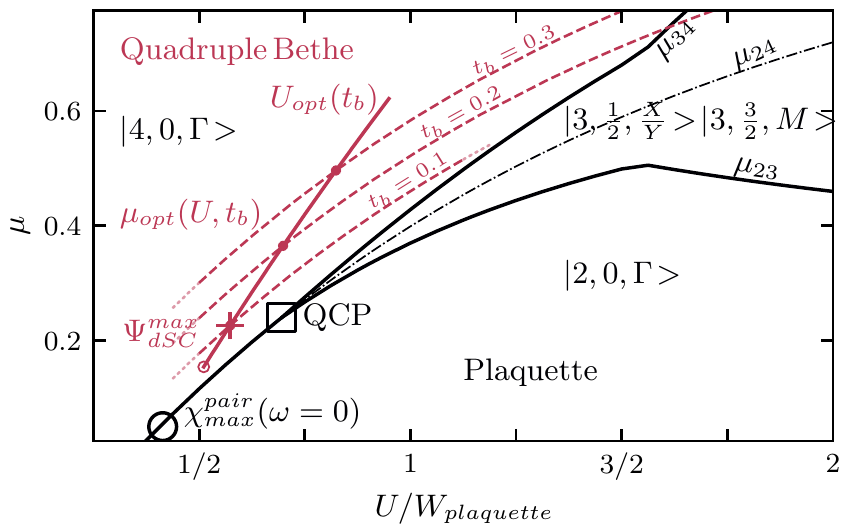}
  \caption{
    Chemical potential $\mu$, Hubbard interaction $U$ -phase diagram of the isolated plaquette (black, solid) with the groundstates $\ket{N,S,K}$ ($t^\prime = 0.3$). $W_{plaquette}$ is the energy range of the plaquette-hopping. The crossover of $\ket{2,0,\Gamma}$ and $\ket{4,0,\Gamma}$ is also shown for $U>U_{QCP}$, where it is not a groundstate crossover (black, dash-dotted). The highly degenerate quantum critical point (QCP) and the maximum of the retarded pairing susceptibility $\chi^{pair}_{max}(\omega = 0)$ of the plaquette are marked. Linear fits of $\mu_{opt}(U, t_b)$ (red, dashed) and $U_{opt}(t_b)$ (red, solid) of the quadruple Bethe lattice are shown. The maximum $\Psi^{max}_{dSC}$ corresponding to the parameter set $(\mu_{opt},U_{opt},t^{opt}_b)$ is marked by $+$.}
  \label{fig:pd_opt}
\end{figure}
In order to sum up the numerical calculations shown in this section we present an overview of the fitted models of the quadruple Bethe lattice's  $\mu_{opt}$ and $U_{opt}$ in the context of the isolated plaquette groundstate phase diagram, see \reffig{fig:pd_opt}. At small $U$ the plaquette exhibits a transition from $\ket{2,0,\Gamma}$ to $\ket{4,0,\Gamma}$ at $\mu_{24}$. For $U>U_c$ this crossover is not a groundstate crossover. However, $\mu_{opt}$ of the quadruple Bethe lattice is parallel to it, indicating that the optimal plaquette state superposition for dSC requires a certain, $t_b$-proportional, gapsize between $\ket{2,0,\Gamma}$ and $\ket{4,0,\Gamma}$. Upon varying $t_b$, $\Psi^{max}_{dSC}$ of the quadruple Bethe lattice stays in the $\mu-U$ diagram closer to the QCP than to the maximum of the pairing susceptibilty of the isolated plaquette.

The quadruple Bethe lattice effectively provides an environment for the states of the isolated plaquette. Neither of the two distinct points, QCP and $\chi^{max}_{dSC}$, in $\mu$-$U$-diagram of the isolated plaquette is the optimal parameter set for the maximum of the dSC order parameter of the quadruple Bethe lattice $\Psi^{max}_{dSC}$. This is due to effective environment shifting dependent on $t_b$ the crucial properties of the QCP, in particular the spectral density peak (\reffig{fig:ak}), to different values of $\mu$ and $U$. The peak at the Fermi level is due to the $N=2,3,4$ degeneracy at the QCP. We will investigate how this feature is related to $\Psi^{max}_{dSC}$ in \refsec{sec:spectral}. The qualitative behavior around the QCP for different $t^\prime$ are very similar despite the QCP being at different ($\mu$, $U$). Thus, at least for small $t_b\sim 0.1$, the dSC properties are governed by the proximity of the QCP. Large $t_b > 0.3$ make the description of the dSC more complicated as transitions between plaquette states other than $\ket{2,0,\Gamma}$, $\ket{3,\frac{1}{2},\frac{X}{Y}}$ and $\ket{4,0,\Gamma}$ become active. Those will also change the optimal doping as shown in \reffig{fig:muoptfitdscdop}. 

In \reffig{fig:pd_opt} we choose to present $U$ with respect to the energy range of the plaquette hopping $W_{plaquette} = 4|t|$. This ratio is interesting in a sense that the square lattice, that is more accurately applied as a description for the cuprates, has a bandwidth of $W_{2d}=8|t|$ and this estimated factor of $2=W_{2d}/W_{plaquette}$ puts our result in a context with $U$-induced correlations of Mott physics studied before with (C)DMFT. With this normalization the QCP lies at $U/W_{plaquette} \approx 0.75$ and the maximum dSC order parameter at $U/W_{plaquette} \approx 0.5$, which can be regarded as intermediate coupling strengths\cite{Comanac2008}.

\section{Component analysis of the Hubbard interaction\label{sec:pkcoupling}}
In \refeq{eq:utransf} we transform the local interaction $U$ into the plaquette-momentum/orbital basis and observe the existence of many two-particle couplings between the plaquette momenta\cite{Merino2014,Gunnarsson2018}, that we classify into intra-orbital repulsion, inter-orbital repulsion, spin-flip,  pair-hop and correlated hopping terms. In this section we investigate the effect of those on the dSC order, but we restrict the discussion to the $X/Y$-orbitals, that are close to Fermi level and describe the dSC order parameter.

\begin{figure}
  \centering
  \begin{tikzpicture}
    \draw (0,0) node{\includegraphics{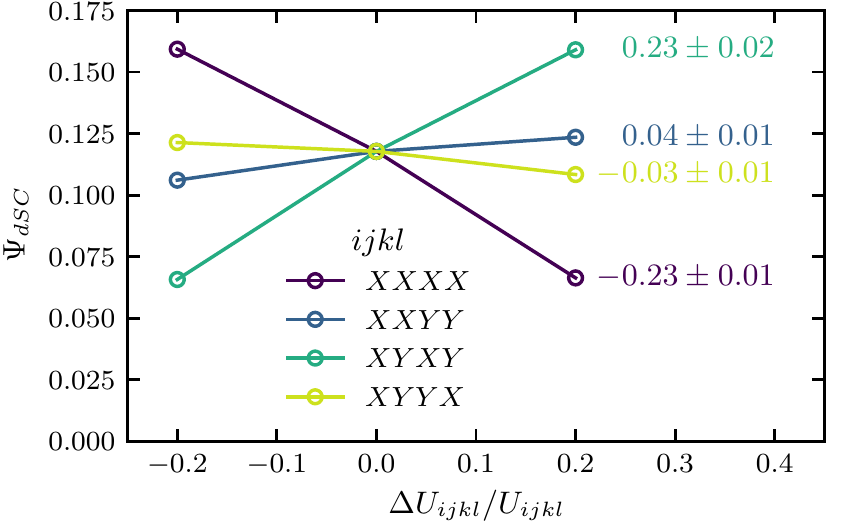}};
    \draw (1.75,-0.55) node{$X$} (3.25,-0.55)node{$Y$}; 
    \draw (1.25,-1) -- (2.25,-1) (2.75,-1) -- (3.75,-1);
    \draw (1.75,-1) node{\large $\uparrow \downarrow$};
    \draw[->] (1.775,-1+0.2) arc[radius=1, start angle = 130, end angle = 50];
    \draw (1.25,-1.5) -- (2.25,-1.5) (2.75,-1.5) -- (3.75,-1.5);
    \draw (1.75,-1.5) node{\large$\uparrow$} (3.25,-1.5) node{\large$\downarrow$};
    \draw[->] (1.95,-1.3) arc[radius=.2, start angle = 90, end angle = -90];
    \draw[<-] (3.05,-1.3) arc[radius=.2, start angle = 90, end angle = 270];
  \end{tikzpicture}
  \caption{$d$-wave superconducting order parameter $\Psi_{dSC}$ as a function of the change in different entries of the Hubbard-interaction $\Delta U_{ijkl}$ normalized by its initial value $U_{ijkl}$ ($t^\prime=0.3$, $t_b = 0.2$, $\delta = 0.15$ and $U=2.78$). Further shown are slopes of linear fits (right) and an illustration (bottom, right) of the two-particle fluctuations, i.e. pair hopping and spin flip.}
  \label{fig:sco_corr}
\end{figure}
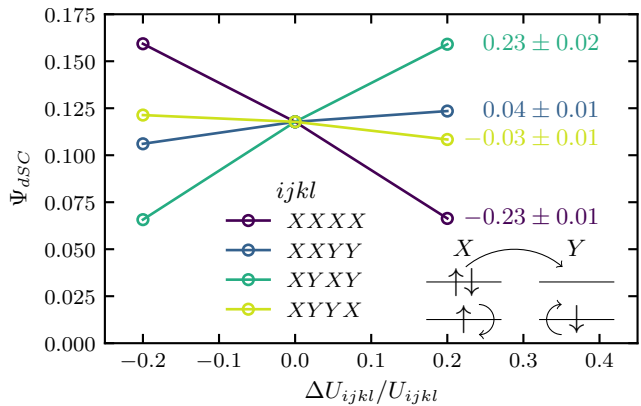
Regarding the notation we introduce the tensor $U_{ijkl}$ for convenience. Initially all of its values are either ``$0$'' or ``$U/4$'', see \refeq{eq:utransf}. In \reffig{fig:sco_corr} we change $U_{ijkl}$ by $20\%$ ($\Delta U_{ijkl}/U_{ijkl} = \pm 0.2$) and observe its effect on the dSC order parameter. Throughout, we change all terms falling into the same class, e.g. a reduction of $U_{XXYY}$ means also a reduction of $U_{YYXX}$. The terms of $U_{ijkl}$, shown in \reffig{fig:sco_corr}, have the same degeneracy. Also, we adjust $\mu$ so that $\delta=0.15$. Changing certain parts of $U_{ijkl}$, we can decrease as well as increase $\Psi_{dSC}$. Whereas pair hoppings ($U_{XYXY}$) and inter-orbital repulsion ($U_{XXYY}$) promote the dSC, spin flips ($U_{XYYX}$) and intra-orbital repulsion ($U_{XXXX}$) diminish it. By the magnitude of the change in $\Psi_{dSC}$, we can identify two competitions in the two-particle processes. First, the pair hopping has the same slope as the negative slope of the intra-orbital repulsion ($U \Delta\Psi_{dSC}/\Delta U \sim 0.23$) and second, the spin flip has the same slope as the negative slope of the inter-orbital repulsion ($U \Delta\Psi_{dSC}/\Delta U \sim 0.04$).

\reffig{fig:sco_corr} shows that at $\delta = 0.15$ the fluctuations are characterized by pair hopping and intra-orbital repulsion rather than spin flips and inter-orbital repulsion. Both competitions occur between a density-density and a fluctuation term. The dominant contribution to the dSC stems from the pair hoppings that compete with the intra-orbital repulsion. The two-particle interaction terms in the plaquette orbital basis reminds of the Kanamori interaction of a multi-orbital atom with peculiar values of the Hund's exchange coupling. Indeed, a supersite formed by only the next-nearest neighbors of the plaquette has been proposed for a unified description of the superconductivity in cuprates and pnictides.\cite{Werner2016}

\section{Spectral properties \& doping dependence\label{sec:spectral}}
The cuprates become superconducting upon doping whereas at half-filling they are insulating. The insulating state is of interest as it can exhibit crucial correlations, but without free charge carriers. The theoretical concepts of the quantum spin liquid and the resonating valence bond state originate from this insulating behavior\cite{Lee2006,Kyung2006a}. At low temperatures this insulator is hidden behind antiferromagnetic ordering. Antiferromagnetic correlations and insulating behavior at half-filling can be explained by the Mott insulator and the DMFT\cite{Georges1996}. The Mott insulator is characterized by a divergence in the mass renormalization of the quasiparticles and has also been suspected to affect the dSC\cite{Sordi2012}. 

\begin{figure}
  \includegraphics{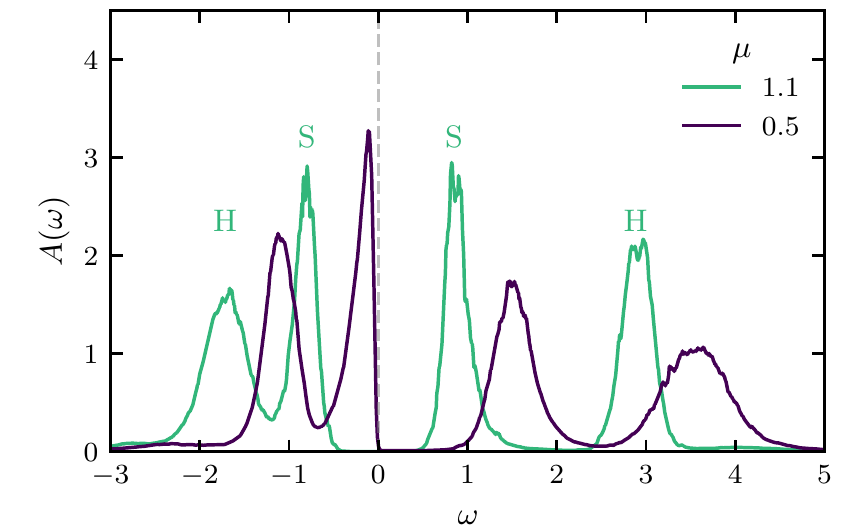}
  \caption{Spectral function $A(\omega)$ for different chemical potentials $\mu$ at approximately half-filling $\delta \approx 0$. For $\mu = 0.5$ we label the one-particle excitations Hubbard (H) and Slater (S) peaks ($t^\prime = 0.3$, $t_b = 0.2$, $U=2.78$).  The analytic continuation is obtained by the stochastic optimization method\cite{Krivenko2019,Mishchenko2000,Pavarini2012}.}
  \label{fig:somins}
\end{figure}
The value of the Hubbard interaction $U$ to model the cuprates is known only approximately\cite{Comanac2008}, and it is debatable whether dSC is a weak- or strong-coupling phenomenon. In \reffig{fig:somins} we present the density of states of the quadruple Bethe lattice at $\delta \approx 0$. It is obtained by the stochastic optimization analytic continuation\cite{Krivenko2019,Mishchenko2000,Pavarini2012} of the (impurity) Green function. At $\mu =1.1$ we observe almost symmetric gap edges formed by two Slater peaks, and with decreasing $\mu$, but still within the gap, so that $\delta\approx 0$, an  asymmetry develops. The hole excitation peak becomes sharper and shifts towards Fermi level. A structure similar to this four-peak structure of two Slater peaks within the Hubbard gap has been found in a prior study for $t^\prime = 0$, and is characteristic of Slater physics that include short-range singlet correlations\cite{Gull2008,Park2008,Go2015}. Correlated singlets also appear in the double Bethe lattice\cite{Hafermann2009,Najera2018}, and define the low-energy excitations at intermediate coupling strengths.

The hole-doped copper-oxide superconductors have a peculiar phase of the pseudogap at underdoping and temperatures above $T_c$. CDMFT studies have shown that its opening can be related to a topological Lifshitz transition at that the Fermi surface turns from electron- to hole-like\cite{Wu2018,Braganca2018}. It can be defined as the point at that the renormalized quasiparticle energy of the $K = X/Y$ points
\begin{equation}
  \label{eq:eqp}
  \tilde{\epsilon}_K = Z_K (t^p_K + \Re \Sigma_K(0))
\end{equation}
cross the Fermi level. $Z_K$ is the quasiparticle residue. The importance of a particle-hole symmetry has also been pointed out in the dSC state\cite{Haule2007}. Particularly for the Bethe lattice model we can also define a renormalized band model for the semicircular density of states\cite{Najera2018}
\begin{equation}
  \label{eq:wqp}
  \tilde{W}_K =  Z_K\, 4 t_b.
\end{equation}
We compare the plaquette momenta of the quadruple Bethe lattice to the high-symmetry points of the Brillouin zone of the square lattice, and thus the Lifshitz transition is defined by $\tilde{\epsilon}_{X/Y}$.

\begin{figure}
  \includegraphics{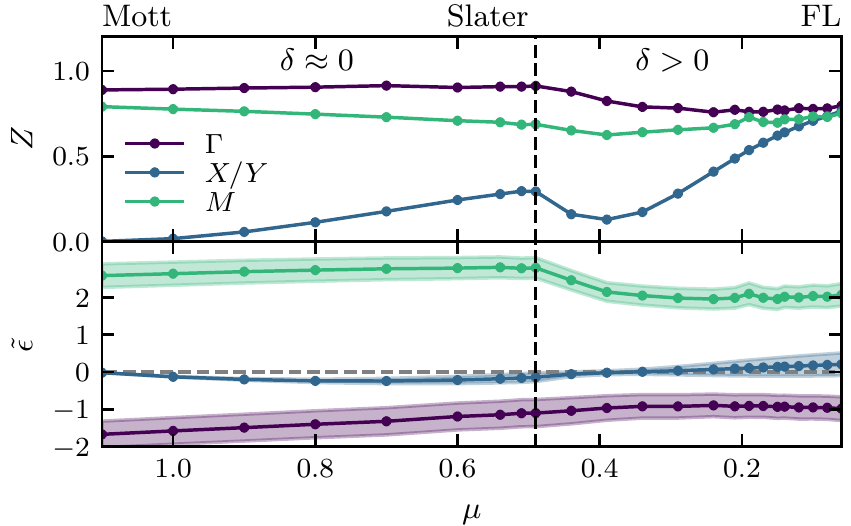}
  \caption{Quasiparticle residue $Z$ (top) and renormalized quasiparticle energy $\tilde{\epsilon}$ and bandwidth $\tilde{W}$ (bottom) of the normal state as functions of the chemical potential $\mu$, that at a certain value (dashed vertical line) hole dopes $\delta$ the system. The $K$-differentiation of $Z$ is absent in the Fermi liquid (FL) ($t^\prime = 0.3$, $t_b = 0.2$ and $U=2.78$).}
  \label{fig:zepsmu}
\end{figure}
\reffig{fig:zepsmu} shows the evolution of the quasiparticle residue $Z$ and the renormalized quasiparticle bands ($\tilde{W}, \tilde{\epsilon}$) with decreasing $\mu$. We use it to continuously tune the insulator into the hole-doped regime. The approximate half-filling region $\delta \approx 0$ on the hole-doped side extends over a large region of $1.1> \mu > 0.5$. The role of $\mu$ is here reminiscent of a field effect transistor experiment in that the spectral properties of the hole excitations change due to the gate voltage.

The Mott phase\cite{Georges1996} is found near $\mu=1.1$, in the center of the gap, where the quasiparticle residue vanishes $Z_X \approx 0$. The system restores coherence in the plaquette orbital $X$ with decreasing $\mu$. The renormalized band model assumes that the self-energy makes only small contributions and renormalizes the quasiparticles of the non-interacting system. For the Mott insulator this assumption is not fulfilled. But for $\mu \lesssim 0.8$ we observe that the renormalized band model agrees with the spectral function from analytic continuation (\reffig{fig:somins}) as both describe the low-energy hole excitation that shifts towards Fermi level. According to the renormalized band model, the spectral properties change and the Mott insulator develops a correlated Slater peak.

In the hole-doped regime $\delta > 0$, we have performed calculations of the normal state for that dSC order is suppressed (\reffig{fig:zepsmu}). Thereby we can investigate quasiparticles and their contribution to the dSC mechanism. $Z_X$ has a local minimum at the Lifshitz transition\cite{Braganca2018,Wu2018}, at that $\tilde{\epsilon}_X = 0$. It is related to a strong scattering rate and suggests an avoided criticality\cite{Haule2007} mechanism of dSC. In the overdoped region the Fermi surface is electron-like and for large hole dopings the plaquette momentum differentiation in $Z_K$ is lifted. In this case, a DMFT description of a (site-)local self-energy can be sufficient for a description, and the system enters the Fermi liquid regime. 

\begin{figure}
  \includegraphics{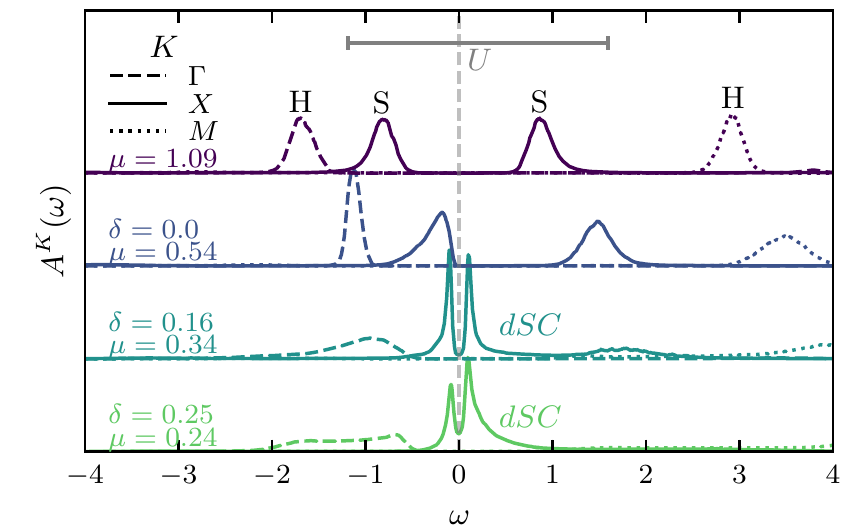}
  \caption{Momentum resolved spectral function $A^K(\omega)$ for different dopings $\delta$ (corresponding to chemical potentials $\mu$) showing a four-peak structure at half-filling ($\delta = 0$) of Hubbard (H) and Slater (S) peaks and for hole-dopings $\delta$ the $d$-wave superconducting gap ($t^\prime = 0.3$, $t_b = 0.2$, $U=2.78$).  The analytic continuation is obtained by the stochastic optimization method\cite{Krivenko2019,Mishchenko2000,Pavarini2012}.}
  \label{fig:som_hssh}
\end{figure}
In \reffig{fig:som_hssh} we show the plaquette-momentum resolved spectral function. It is obtained by analytic continuation of the (local lattice) Green function and shown also for the symmetry-broken dSC state. The Slater peaks describe excitations with momentum $X/Y$. The splitting of upper peaks and lower peaks is of the order of $U$. Decreasing the chemical potential shifts the lower Slater peak to Fermi level, and the dSC order originates from the lower Slater peak, i.e. the dSC gap appears with the doping of the Slater peak, see \reffig{fig:som_hssh}. The spectral function of \reffig{fig:som_hssh} looks very similar to \reffig{fig:ak}, so that it is possible to relate the plaquette transitions $\ket{4,0,\Gamma} \rightarrow \ket{3,\frac{1}{2}, \frac{X}{Y}}$ and $\ket{4,0,\Gamma} \rightarrow \ket{3,\frac{1}{2}, \Gamma}$ to the lower Slater and Hubbard peaks, respectively. It points out the crucial part of $\ket{3,\frac{1}{2}, \frac{X}{Y}}$, that provides low-energy transitions for the electrons that will form the dSC pairs. Further does $\ket{3,\frac{1}{2}, \frac{X}{Y}}$ provide a single-particle transition to $\ket{2,0,\Gamma}$ and the pairs are formed by the latter and $\ket{4,0,\Gamma}$ (\reffig{fig:chixxpair}).

\begin{figure}
  \includegraphics{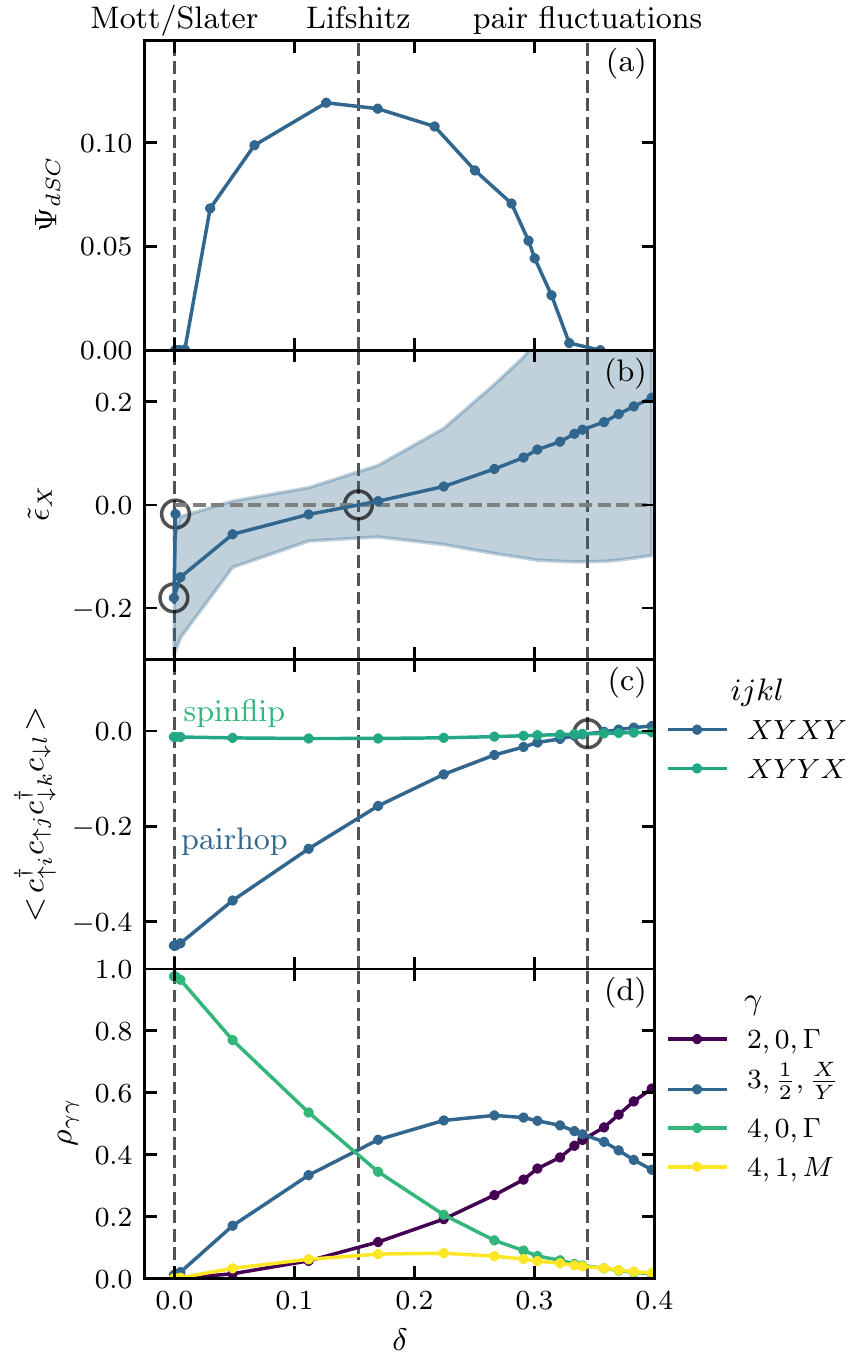}
  \caption{$d$-wave superconducting order parameter $\Psi_{dSC}$ of the symmetry-broken state (a), renormalized quasiparticle band(width) $\tilde{\epsilon}_X$ of the normal state (b), static two-particle Observable $\expval{...}$ of the normal state (c) and reduced density matrix of the normal state $\rho$ with plaquette many-body state indices $\gamma$ (d) as functions of hole-doping $\delta$. Points of certain features in the doping dependence are marked by circles. $t^\prime = 0.3$, $t_b = 0.2$, $U=2.78$.}
  \label{fig:rhoqpmu}
\end{figure}
\reffig{fig:rhoqpmu}~(a) shows the doping dependence of the dSC order for $t_b=0.2$. We characterize the maximum and the endpoints by features in the correlation functions of the normal state with suppressed dSC order (\reffig{fig:rhoqpmu}~(b)-(d)). At half-filling we find the two different solutions of insulators as discussed in \reffig{fig:zepsmu}. In the underdoped regime the single-particle excitations are hole-like and the quasiparticle bandwidth $W_X$ is strongly renormalized. The renormalization is particularly strong at the Lifshitz transition at that the Fermi surface changes from hole-like to particle-like. At this point is also the maximum of the dSC dome. At overdoping the quasiparticle of the $X$-orbital becomes more Fermi liquid-like and the quasiparticle energy shifts away from Fermi level. The renormalized bandwidth broadens and quasiparticle states remain at Fermi level at the overdoping end of the dSC dome.

In the overdoped regime the dynamics of the single-particle correlations do not show any peculiar feature. In order to understand this regime better we present the static two-particle correlation functions in \reffig{fig:rhoqpmu}~(c). In \refeq{eq:utransf} we have discussed the transformation of the local Coulomb repulsion into plaquette momentum basis. The fluctuation terms between $X$ and $Y$ only, i.e. pair hopping and spinflip terms, appear symmetrically in the interaction, but the dependence of $\Psi_{dSC}$ is stronger on the pair hopping part of the interaction, see \refsec{sec:pkcoupling}. At the overdoping end of the dSC dome pair hopping and spin flip correlations between $X$ and $Y$ are equally weak.

Due to the small Bethe hopping (hybridization) the dSC phase is mostly governed by a few low-energy cluster eigenstates. \reffig{fig:rhoqpmu}~(d) shows that the dSC order occurs only where the Boltzmann weights of $\ket{3,\frac{1}{2},\frac{X}{Y}}$ and $\ket{2,0,\Gamma}$ are non-zero. The large pair hopping correlations stem mostly from $\ket{4,0,\Gamma}$. Only a combination of both, $\ket{3,\frac{1}{2},\frac{X}{Y}}$ which produces a peak at Fermi level and pair hopping correlations, results in the non-trivial dome-like structure of the dSC order. At the overdoping end of the dSC dome the Boltzmann weight of the spin-triplet $\ket{4, 1, M}$ exceeds that of $\ket{4,0,\Gamma}$, and pair hoppings correl vanish which suppresses the dSC.

\begin{figure}
  \centering
  \includegraphics{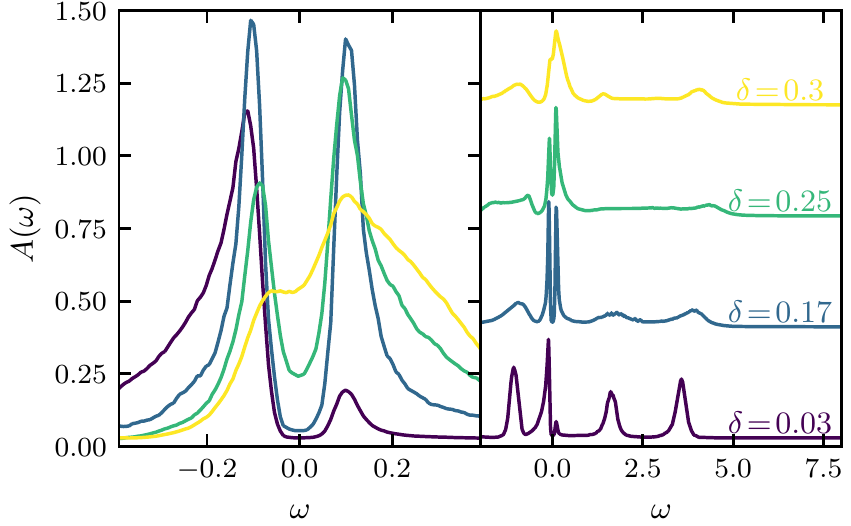}
  \caption{Local density of states for different hole-dopings $\delta$ in the dSC state ($t^\prime = 0.3$, $t_b = 0.2$, $U=2.78$). The color code in the zoom-in (left) is the same as in the overview (right). The analytic continuation is obtained by the stochastic optimization method\cite{Krivenko2019,Mishchenko2000,Pavarini2012}.}
  \label{fig:dscgap}
\end{figure}
\reffig{fig:dscgap} is a detailed view of \reffig{fig:som_hssh} with more values for $\delta$. It shows the dSC gap of the one-particle spectral function. Finite hole doping and dSC order set in with a sharp peak below Fermi level and a small peak above. The latter grows until at optimal doping the dSC gap is approximately particle-hole symmetric. From optimal doping to overdoping the peak of hole excitations shifts through the Fermi level increasing spectral weight at Fermi level until the gap is closed and dSC order is absent. In contrast to the lower edge the upper edge of the dSC gap does not shift with doping. It suggests that two distinct mechanisms contribute to the formation of the dSC gap in the one-particle spectral function\cite{Civelli2008}.

\section{Superconductivity \& Antiferromagnetism\label{sec:otherorders}}
\begin{figure}
  \includegraphics{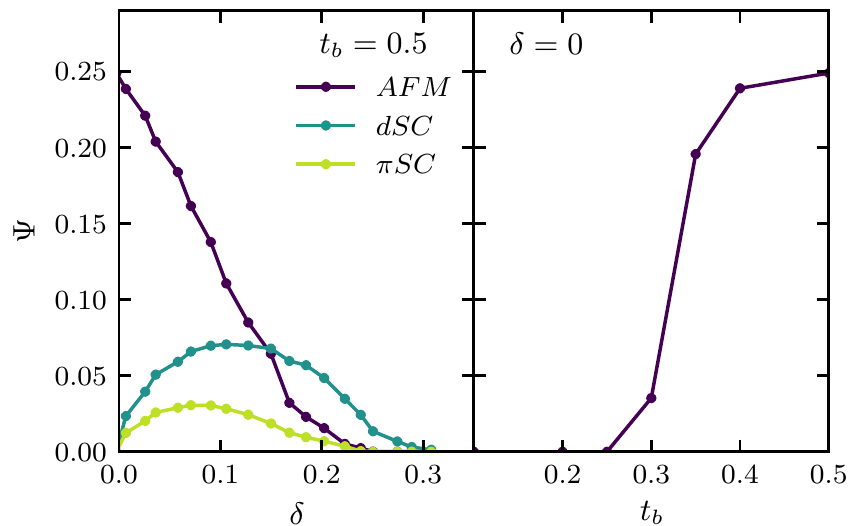}
  \caption{Order parameters $\Psi$ of antiferromagnetism (AFM), $d$-wave superconductivity (dSC) and spin-triplet superconductivity ($\pi SC$) dependent on hole doping $\delta$ for Bethe hopping $t_b=0.5$ (left) and dependent on $t_b$ for half-filling $\delta = 0$ (right) ($U=2.78$, $t^\prime = 0.3$).}
  \label{fig:dscafm}
\end{figure}

Using a two-by-two plaquette as cluster, we can describe AFM and dSC order on equal footing, and both are relevant for the phase diagram of the cuprates. In \reffig{fig:dscafm} we observe that it is largest at half-filling, at that according to experimental findings, the Néel temperature is also largest. In contrast to the hole doped cuprates we find coexistence\cite{Demler2004,Almeida2017} of AFM and dSC order up to $\delta = 0.25$ which is a well-known feature of CDMFT\cite{Lichtenstein2000,Capone2006,Kancharla2008,Foley2018} and is expected to arise from the neglecting of long-ranged correlations. In fact, already an eight-site cluster can suppress dSC in proximity of half-filling\cite{Gull2013}.

The maximum value of the AFM order parameter is $\Psi_{AFM}=0.25$, which is only half the magnitude of the plaquette's full magnetization. This is the case, because two electrons are locked in the singlet of the $\Gamma$-orbital, that is fully occupied and doesn't touch the Fermi-level, see \reffig{fig:nonint}. Finite values for $\Psi_{AFM}$ we find only for $t_b \geq 0.3$, it increases sharply as function of $t_b$ and saturates around $t_b=0.4$ at $\Psi_{AFM} = 0.25$. This is very different from the dSC order parameter, that has its maximum around $t_b = 0.1$ (\reffig{fig:pd_opt}). It seems, that for the AFM it is necessary to have a certain minimal spin exchange interaction not only within but also between plaquettes. In contrast, dSC requires a certain plaquette eigenstate configuration and a much weaker plaquette hybridization. To some extend this asymmetry can be understood regarding the non-interacting density of states (\reffig{fig:nonint} and \refeq{eq:gblocks}). While dSC occurs entirely within $X$ and $Y$, AFM order couples also $\Gamma$ and $M$, which are split and farther from the Fermi level.

Moreover, we observe spin-triplet superconductivity\cite{Chen2015,Foley2018} ($\pi$SC) with the order parameter
\begin{align}
  \label{eq:psipisc}
  \Psi_{\pi SC} =& \frac{1}{4^2} \sum_{R R^\prime} \left( \cos \left[ \pter{X} \left( \pter{R}- \pter{R}^\prime\right)\right] - \cos \left[ \pter{Y} \left( \pter{R} - \pter{R}^\prime\right)\right]  \right)\nonumber \\
                 &\times e^{i M R^\prime} \expval{c_{R\up}c_{R^\prime\dn} + c_{R\dn}c_{R^\prime\up}}.
\end{align}
It is described by entries of the correlation functions that are off-diagonal in Nambu and plaquette-momentum space, see \refeq{eq:gblocks}. Further, a comparison with \refeq{eq:orderparameters} shows also that it is a combination of AFM and dSC as it breaks the spatial symmetries of the plaquette according to AFM and dSC. Finally, the symmetry upon spin-exchange can be seen explicitly in \refeq{eq:psipisc} and stresses the spin-triplet character. We find non-zero values for $\Psi_{\pi SC}$ only at dopings for the the quadruple Bethe lattice also shows coexistence of dSC and AFM. Thus $\pi SC$ is a result of the interplay between dSC and AFM.

\section{Extended Bethe lattice hopping\label{sec:tbnnn}}
To this point the Bethe hopping exists only within one Bethe lattice and is represented by a scalar. In this section we introduce the Bethe-hopping matrix in plaquette-site space
\begin{equation}
  \label{eq:matrixtb}
  \underline{t_b} =
  \begin{pmatrix}
    t_b & 0 & 0 & t^\prime_b\\
    0 & t_b & t^\prime_b & 0\\
    0 & t^\prime_b & t_b & 0\\
    t^\prime_b& 0 & 0 & t_b
  \end{pmatrix}
\end{equation}
with the extended Bethe-lattice hopping $t^\prime_b$, that appears in entries, that in the case of the plaquette hopping matrix $t_p$ are occupied by the next-nearest neighbor hopping. It means, that for $t^\prime_b$ the transition between plaquettes is accompanied by a transition to the next-nearest neighbor of the target plaquette. The effect of the non-diagonal terms in the $\underline{t}_b$-matrix is the finetuning of the widths of the semicirculars independently. The nearest neighbor components would affect only the widths of the $\Gamma$ and $M$ bands, so we have set them to zero for simplicity. 

The CDMFT self-consistency becomes
\begin{equation}
  \label{eq:selfconsistencytbmat}
  G^{-1}(i\omega_n)=\left(i\omega_n+\mu\right) \mathbbm{1} -t^{p} -\underline{t_{b}}G(i\omega_n)\underline{t_{b}}
\end{equation}
with the quantities being matrices in plaquette site space (Nambu degrees of freedom are omitted for convenience), and the last term are matrix products of $\underline{t_b}$ and $G$.

\begin{figure}
  \centering
  \includegraphics{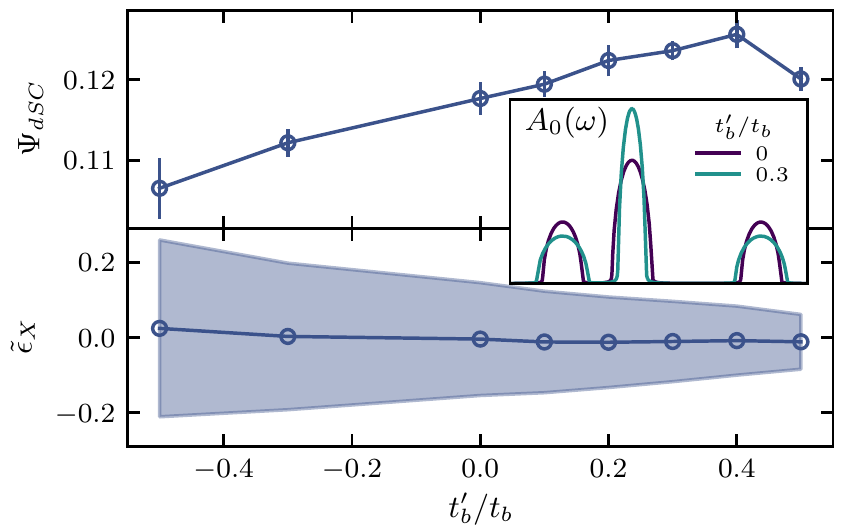}
  \caption{Superconducting order parameter $\Psi_{dSC}$ (top) and renormalized quasiparticle energy $\tilde{\epsilon}_X$ (bottom) as functions of the extended Bethe lattice hopping at $\delta = 0.15$ ($U = 2.78$, $t^\prime = 0.3$, $t_b = 0.2$). The colored area marks the renormalized quasiparticle bandwidth $\tilde{W}$. Non-Interacting semicircular density of states for different next-nearest neighbor Bethe lattice hoppings (inset).}
  \label{fig:scotbp}
\end{figure}
The transformation \refeq{eq:ktransform} must be applied to $\underline{t_b}$, too. Thus it gets diagonalized. The non-interacting semicircular density of states changes so that the width and height of $G$, $M$ differ from those of $X$, $Y$, but they remain semicirculars, see \reffig{fig:scotbp} (inset). Depending on the sign of $t^{\prime}_b/t_b$ the height increases and the width decreases or vice versa. Despite the absence of a real divergence the effect of an increased density of states of $X$ and $Y$ can be interesting in the context of the van Hove singularity\cite{Markiewicz1997,Piriou2011} in the square lattice.

\reffig{fig:scotbp} shows that the dSC order parameter increases with a decreasing quasiparticle bandwidth. The doping is set to $\delta = 0.15$ that remains independent of $t^\prime_b$ related to the Lifshitz transition, i.e. the quasiparticle energy $\tilde{\epsilon}_X$ is almost constant. The change in $\Psi_{dSC}$ is small, and therefore in \reffig{fig:scotbp} we also present errorbars, that are calculated as the largest absolute deviation of eight values of two CDMFT-loops with local Green functions that by symmetry have four entries of the order parameter. The quasiparticle residue of the presented calulcations only weekly depends on $t^\prime_b$, i.e. stays $Z_X=0.38\pm0.02$. Therefore the change of the quasipartilce's bandwidth with $t_b^\prime$ is almost entirely due to the renormalization of the non-interacting $X/Y$-band. 

The modification of the semicircular density of states by $t^\prime_b$ is small and finite, but it already shows an enhancing effect on $\Psi_{dSC}$. The van Hove singularity corresponds to an infinite density of states and can potentially enhance that effect much more. The role of the next-nearest neighbor hopping $t^\prime$ in the cuprates is cumbersome. Whereas a comparison of bandstrucutre calculations with experiments show a finite value for $t^\prime$ as the optimal one\cite{Pavarini2001}, calculations in the framework of strong correlations are not able to confirm this by including only local correlations. First, we observe that $t^\prime$ shifts the quantum critical point of the plaquette, which in the quadruple Bethe lattice is in proximity to the maximum $\Psi_{dSC}$, to smaller $U$. And second, we find that a hopping similar to $t^\prime$, i.e. $t^\prime_b$, can have an enhancing effect on $\Psi_{dSC}$. It reduces effectively the bandwidth of $X$ towards the optimal value $t_b\approx 0.1$ (\reffig{fig:uoptscoopttb}). Finally, it is important to note that the diminishing effect of the next-nearest neighbor hopping in CDMFT studies of the square lattice is not necessarily a contradiction with bandstructure calculations on cuprates, since it may also indirectly support dSC by suppressing other, competing, orders e.g. stripes\cite{Jiang2018}.

\section{Conclusion}
We have formulated the CDMFT self-consistency, \refeq{eq:selfconsistency}, that solves the quadruple Bethe lattice exactly, also in the $d$-wave superconducting state. An analysis of the isolated two-by-two cluster has shown that this plaquette is even without an environment unstable towards dSC order. The coupling to other plaquettes in the infinite dimensional quadruple Bethe lattice allows for the spontaneous symmetry breaking. dSC order is found in proximity of a QCP of the plaquette in the $\mu$-$U$ diagram, where the plaquette eigenstates $\ket{2,0,\Gamma}$, $\ket{3,\frac{1}{2},\frac{X}{Y}}$ and $\ket{4,0,\Gamma}$ cross. The optimal value for the parameter that controls the hybridization of plaquettes is $t_b=0.1$ with the optimal doping $\delta = 0.15$. The latter has also been measured in experiments on the cuprates.

The dSC dome of the doping phase diagram lies next to a half-filling state with a vanishing quasiparticle residue, characteristic of the Mott insulator. Moreover, at half-filling the renormalized quasiparticle picture shows a crossover to an insulator with a correlated Slater peak with decreasing $\mu$. The hole excitations corresponding to the Slater peak occur around the energy of the $\ket{4,0,\Gamma} \rightarrow \ket{3,\frac{1}{2},\frac{X}{Y}}$-transition of the isolated plaquette. At hole doping this hole excitation forms the superconducting gap. The small density of states at Fermi level restricts the local pair formation in the underdoped regime, a Lifshitz transition occurs at optimal doping, and at overdoping the superconductivity is suppressed by the vanishing of the two-particle pair-hopping correlations.

For large $t_b = 0.5$ the model exhibits AFM that coexists with dSC order. Since the AFM does not exist at $t_b=0.1$ this additional parameter allows for a disentanglement of these two orders. It can be used to model effects beyond the cluster effectively, e.g. by an increase of the non-interacting density of states at Fermi level, reminiscent of a van Hove singularity. The latter can enhance the dSC. Additionally, in the coexistence regime of dSC and AFM exists a spin-triplet type of superconductivity, $\pi$SC. Whereas AFM is staggered within each of the four Bethe lattices of the quadruple Bethe lattice, dSC and $\pi$SC are homogenous in those.

\begin{acknowledgments}
We thank A.~Georges and A.~Millis for discussions. MH, SB and AIL acknowledge support by the Cluster of Excellence 'Advanced Imaging of Matter' of the Deutsche Forschungsgemeinschaft (DFG) - EXC 2056 - project ID 390715994 and by the DFG SFB 925. MIK acknowledges financial support from NWO via Spinoza Prize. The computations were performed with resources provided by the North-German Supercomputing Alliance (HLRN).
\end{acknowledgments}

\bibliography{quadruplebethe}

\end{document}